\documentclass{emulateapj}
\shortauthors{Toft et al.}
\shorttitle{The Dynamical Mass-Size relation at  $z\sim2$, and First
  Constraints on the Fundamental Plane}
\usepackage{amsbsy} 
\usepackage{amsfonts} 

\begin{document}

\title{ Deep Absorption Line Studies of Quiescent Galaxies at
  $z\sim2$: The Dynamical Mass-Size Relation, and First Constraints on
  the Fundamental plane\footnote{Based on X-Shooter-VLT observations
    collected at the European Southern Observatory, Paranal, Chile
    (program IDs 084.A-0303, 084.A-035)}}

\author{S. Toft\altaffilmark{1},  A. Gallazzi\altaffilmark{1},
  A. Zirm\altaffilmark{1} , M. Wold\altaffilmark{1}, S. Zibetti\altaffilmark{1,2},
  C. Grillo\altaffilmark{1}, A. Man\altaffilmark{1}}

\altaffiltext{1}
{Dark Cosmology Centre, Niels Bohr Institute, University of Copenhagen, Juliane Mariesvej 30, DK-2100 Copenhagen, Denmark, sune@dark-cosmology.dk}
\altaffiltext{2}
{INAF - Osservatorio Astrofisico di Arcetri, Largo Enrico Fermi 5, I-50125 Firenze, Italy}

\begin{abstract}
  We present dynamical and structural scaling
  relations of quiescent galaxies at $z=2$, including 
  the dynamical mass-size relation and the first constraints on the fundamental plane
  (FP).  The backbone of the analysis is a new, very deep VLT/X-shooter spectrum
  of a massive, compact, quiescent galaxy at $z=2.0389$. We detect the
  continuum between 3700-22000{\AA} and several strong
  absorption features (Balmer series, Ca H+K, G-band), from which we derive a
  stellar velocity dispersion of $318\pm53$ km s$^{-1}$.  We perform
  detailed modeling of the continuum emission and
  line indices and derive strong simultaneous constraints on the age,
  metallicity, and stellar mass.  The galaxy is a dusty
  ($A_V=0.77^{+0.36}_{-0.32}$)  solar metallicity (${\rm log}(Z/Z_{\odot}) =
  0.02^{+0.20}_{-0.41}$) post starburst galaxy, with a mean luminosity weighted log(age/yr)
  of $8.9\pm0.1$. The galaxy formed the
  majority of its stars at $z>3$ and currently has little or no ongoing star
  formation. We compile a sample of three other $z\sim2$ quiescent
  galaxies with measured velocity dispersions, two of which are also
  post starburst like. Their dynamical mass-size relation is offset significantly less
 than the stellar mass-size relation from the local
  early type relations, which we attribute to a lower central dark
  matter fraction. Recent cosmological merger simulations qualitatively agree with the
data, but can not fully account for the evolution in the dark matter fraction.
The $z \sim 2$  FP requires additional
evolution beyond passive stellar aging, to be in agreement with the
local FP. The structural evolution predicted by the
 cosmological simulations is insufficient, suggesting that additional, possibly non-homologous structural evolution is needed.
\end{abstract}

\section{Introduction}

Since first being discovered in NIR surveys \citep[e.g.,
FIRES][]{franx2003}, massive, evolved galaxies at $z\sim2$ have been a
subject of extensive research, characterizing their properties
from increasingly larger and more complete samples \citep[e.g.][]{
  vandokkum2006,franx2008, toft2009, mccracken2010, williams2010,
  newman2011, damjanov2011, cassata2011, wuyts2011}.  By combining photometric
redshifts, stellar masses, and star formation histories derived from
Spectral Energy Distribution (SED) fits to broad band photometry
investigators have probed their general properties: They are massive
(M$>$10$^{11}$ M$_{\odot}$) galaxies, with space densities $5\times10^{-4} $Mpc$^{-3}$
\citep{vandokkum2010, marchesini2009} which cluster strongly
\citep{daddi2003, quadri2008} and approximately $30-50\%$ of them are
quiescent, with little or no star formation
\citep[e.g.][]{kriek2006, toft2007, toft2009, kriek2009}. In high
resolution imaging, the star forming galaxies have extended and in
some cases disturbed morphologies, while the quiescent galaxies are
uniformly extremely compact (effective radius $r_e\lesssim$ 1kpc),
corresponding to average stellar mass densities (within $r_{e}$) 1--2
orders of magnitudes higher than in local early types
\citep{daddi2005, toft2005, trujillo2006, zirm2007, toft2007,
  cimatti2008, kriek2009, conselice2011,cassata2011, wuyts2011}, but see also
\cite{mancini2010, newman2010}. Furthermore, there is evidence that a
significant fraction of $z\sim2$ quiescent galaxies have flat, exponential disk
like surface brightness profiles, rather than the $r^{1/4}$ profiles
found in local quiescent galaxies \citep{toft2005, toft2007,
  vandokkum2008, vanderwel2011, wuyts2011}

If galaxies with these properties exist in the local universe, they
are extremely rare \citep{trujillo2009, taylor2010, shih2011}.  The
evolutionary path of the compact quiescent galaxies to the local
universe is not well understood. Their high stellar masses, compact
morphologies and quiescent nature makes it natural to assume they are
the progenitors, or ``seeds'' of massive local early type galaxies,
which is also supported by number density arguments
\citep{vandokkum2010} but they have to go through significant
structural evolution, increasing
their effective radii by factors of 3-6, and possibly transforming their flat exponential disk like
structure into bulge-dominated/early type systems.
 In the following we will
refer to these massive quiescent compact $z\sim2$ galaxies as
SEEDs. Minor and major dry merging are the primary candidate processes
for puffing up their sizes and transforming their profiles from flat
disk-like systems to spheroidal early types
\citep[e.g.][]{khockfar2006, naab2009, bezanson2009}.
  
One of the main uncertainties in the interpretation of the properties of
SEEDs is that it is mainly based on photometric redshifts and SED fits
to broadband photometry.  Due to their quiescence, they are void of
strong emission lines and have very faint rest-frame ultraviolet (UV)
continua. It is challenging to detect absorption lines, that can
be used for redshift determination of quiescent systems.  The red rest-frame optical color of SEEDs makes them
brighter in the observed near-infrared (NIR), but due to the limited
sensitivity of NIR spectrographs, spectroscopic confirmation have only
been possible for small samples of the very brightest examples. In a
NIR spectroscopic survey of nine of the brightest examples, using the
Gemini Near Infrared Spectrograph (GNIRS), \cite{kriek2006} were able
to detect the continuum well enough to determine the redshift from the
position of the 4000 {\AA} break, and confirm the relatively old age
(1-2 Gyr) and quiescence from the shape of the continuum and lack of
strong emission lines.  Even with relatively long exposure times (1-4
hours) on an 8m telescope it was not possible to make significant
detections of absorption lines. Only in a much deeper (29 hour)
follow-up spectrum were absorption lines detected for one
object \citep{kriek2009}.  In this case, it was also possible to
measure the stellar velocity dispersion of a $z\sim2$ galaxy for the
first time.  The dispersion is very high, consistent with what was
expected from its high derived stellar mass density \citep{toft2007,
  cimatti2008,vandokkum2009}. Since then velocity dispersions have
been derived for a few handful of quiescent galaxies in the redshift range
$1<z<2$ \citep{cenarro2009, cappellari2009, newman2010, onodera2010, vandesande2011}.

With the advent of new, more sensitive spectrographs it is now
feasible to obtain ``absorption line quality'' spectra for samples of
SEED galaxies in reasonable amounts of time.  In this paper we present
the first results of such a survey that we conducted with the
X-shooter spectrograph on the Very Large Telescope. As this is the
first detailed spectroscopic investigation of this type of galaxy, we
describe stellar population and velocity dispersion fits in
considerable detail.

The structure of the paper is as follows. In Section \ref{sec.observations} we
describe the target selection, acquisition and reduction of a deep
X-shooter spectrum of a quiescent galaxy at $z\sim2$. In Section \ref{sec.spectralanalysis} and \ref{sec.stellarpops} we perform a
detailed analysis of the spectrum, fitting the UV-NIR continuum and
absorption lines indices, with stellar population synthesis models, to
derive stringent constraints on the age, stellar mass, dust
content and metallicity of the galaxy. In
Section \ref{sec.surfacebrightnessfits} we fit the surface brightness
profile of the galaxy to derive its structural parameters. In
Section \ref{sec.velocitydispersion} we measure the velocity dispersion
from the absorption lines, and derive the dynamical mass. In
Section \ref{sec.sample} we compile a sample of 3 other $z\sim2$ galaxies
from the literature with measured velocity dispersions, and compare
scaling relations between their structural and dynamical properties
with those observed in local galaxies, to derive constraints
on the dynamical-mass size relation and, for the first time,  the fundamental plane relations at z=2.
Finally we summarize and discuss the
results in Section \ref{sec.discussion}. Throughout the paper we assume a
standard flat cosmology ($\Omega_{\Lambda},\Omega_{M}=(0.7,0.3),
    H_0=73$ km s$^{-1}$Mpc$^{-1}$). All magnitudes are referenced to
    the AB system \citep{oke1974}

\begin{center} 
\begin{deluxetable*}{l l l}
\small
\tablecaption{ Details of the X-shooter observations}
\tablecolumns{3}
\tablewidth{12cm}
\tablehead{ \colhead{Night} & \colhead{Exptime [s]} & \colhead{Slit size[$\arcsec$]} \\ \colhead{} & \colhead{[UVB/OPT/NIR]} & \colhead{[UVB/OPT/NIR]} } 
\startdata
10/19/2010 &  $4\times800/4\times800$/$4\times900$ &
$0.8\times11/0.9\times11/0.6\times11$  \\ 
10/20/2010 & $8\times800/8\times800/8\times 900$ &
$1.0\times11/0.9\times11/0.9\times11$  \\
10/21/2010 & $8\times800/8\times800/8\times 900$ &
$1.0\times11/0.9\times11/0.9\times11$  \\
\enddata
\label{tab.observations}
\end{deluxetable*}
\end{center}

\section{Observations}
\label{sec.observations}

\subsection{Target selection}
Our objective was to take advantage of the unique combination of
sensitivity, spectral coverage and resolution of X-shooter to obtain a
spectrum of a SEED galaxy of unprecedented quality, to allow a
detailed study of the absorption lines for the first time, and to
obtain a spectrum of sufficient S/N to be used as a template for
designing programs to observe larger samples.
 
\begin{figure}
\begin{center}\includegraphics[scale=0.47,angle=0]{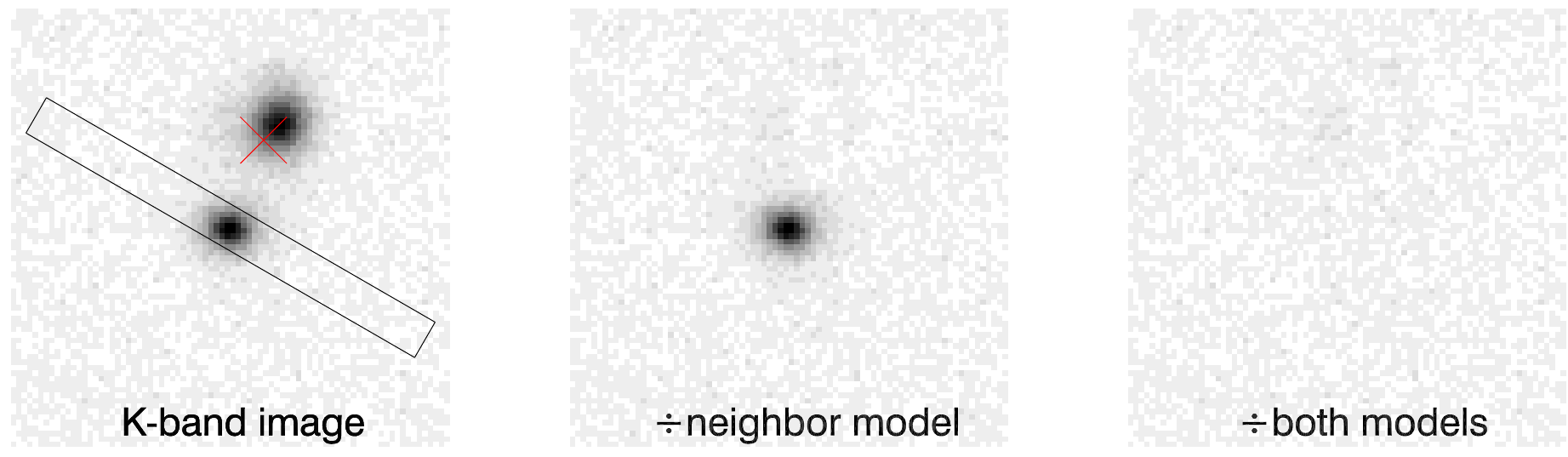}
  \caption{{\bf Left}: K-band image ( $10.7\arcsec \times
    10.7\arcsec$) of UDS19627 (WFCAM), with the position of the
    X-shooter $11\arcsec \times 0.9\arcsec$ slit indicated. North is
    up, East is to the left. The seeing is $0.7\arcsec$. The brighter
    galaxy north-west of the target is at lower redshift.  {\bf
      Middle}: We modelled the surface brightness profiles of the two
    galaxies simultaneously with {\sc galfit}. In this image the best
    fitting model of the neighbor has been subtracted, demonstrating
    its minimal influence on our fit of the main target. {\bf Right:}
    Residual after subtracting both best fitting surface brightness
    models.  The red cross indicates the centroid of a $24\micron$
    source associated with the foreground galaxy.}
\label{fig.images}
\end{center}
\end{figure}

The target UDS19627 (RA=02:18:17.1, DEC=-05:21:38.8 ) was chosen from
the UKIRT Ultra Deep Survey (UDS), as a (K-band) bright example of a compact, quiescent, massive galaxy with
$z\sim2$ \citep{williams2010}. 
Based on the broad band photometry (See Table \ref{tab.photometry}), the
target is a very massive, quiescent galaxy at
$z_{phot}=2.02_{-0.08}^{+0.07}$ with K=20.19$\pm 0.01$.
Images of the target are shown in Figure \ref{fig.images}. The bright galaxy $\sim
3\arcsec$ north-west of the target is at lower redshift
\citep[UDS19771, $z_{phot}=0.58$,][]{williams2010}. This galaxy may be boosting the brightness of our target slightly, through
the gravitational lensing effect (see Section \ref{sec:caveats})

\begin{deluxetable*}{ l l l l l l l l l l l}
\label{tab.photometry}
\tablecaption{Photometry: AB magnitudes for UDS19627 from the UDS
  survey \citep{williams2010}. Ch1 and Ch2 refer to Spitzer/IRAC 3.6 and
  4.5{\micron} bands}
\tablecolumns{11}
\tablewidth{0pt}
\tablehead{
\colhead{U} & \colhead{B} & \colhead{V} & \colhead{R} & \colhead{I} &
\colhead{z} & \colhead{J} & \colhead{H} & \colhead{K} & \colhead{Ch1}&
\colhead{Ch2}  }
\startdata
25.72 & 24.76 & 23.94 & 23.47 &
23.15 & 22.62 & 20.94 & 20.46 &
20.19 & 19.67 & 19.58 \\
$\pm$0.21 & $\pm$0.03 & $\pm$0.03 & $\pm$0.03 &
$\pm$0.03 & $\pm$0.03 & $\pm$0.03 & $\pm$0.03 &
$\pm$0.02 & $\pm$0.03 & $\pm$0.04
\enddata
\end{deluxetable*}

\subsection{X-shooter Data}
We obtained deep spectra with the X-shooter spectrograph, mounted on
ESO's Very Large Telescope. X-shooter is a single object echelle
spectrograph, consisting of three arms which cover simultaneously the
spectral range 3000-25000{\AA} \citep{vernet2011}.  We obtained a
total exposure time of five hours over three nights from October 19-21
2009, as part of the X-Shooter GTO program. Details of the
observations are given in Table \ref{tab.observations}. On the first
night the seeing was very good (FWHM$\sim 0.5\arcsec$), so we used a
narrow slit. On the following nights the seeing was
$0.7\arcsec-1.0\arcsec$, so we switched to a larger slit width, to
minimize slit loss.  We performed a blind offset to the object from a
nearby bright star, and nodded on the slit with a ABA$^\prime$B$^\prime$ sequence with
4 independent positions.

\subsection{Reduction}
We reduced the data using a combination of the ESO X-shooter
pipeline (version 1.2.0 in physical mode) and our own scripts.
We experimented with different reduction recipes. 
The cleanest, highest S/N spectrum (between the skylines)  was achieved when reducing the
individual exposures in STARE mode. 
To determine accurate offset between the trace in different exposures,
we ``collapsed'' regions between skylines, in the rectified, sky
subtracted frames and measured the shifts between the peaks of the
spectral point spread functions (SPSF, the object is effectively a
point source due to the seeing).  We then shifted the individual
exposures to the same position and combined them by computing the
sigma clipped average of each pixel (masking out bad pixels and cosmic
rays).  We also created a 2D variance map in which each pixel is the
variance of the 20 pixels in the stack.

The individual background subtracted, rectified 2D NIR frames were
corrected for telluric absorption, using telluric star observations
that were observed immediately before or after each 1 hour observation
block.

This correction was then applied to each row in the individual 2D
frames before they were combined. Finally, we extracted the 1D
spectrum, using optimal extraction \citep{horne1986}, weighing the
pixels both by the SPSF and by their variance.  For flux calibration
we used a spectrum of the spectro-photometric standard standard BD
+17$^{\circ}\mathrm{C}$4708 \citep{bohlin2004} in combination with the
CALSPEC HST database \citep{bohlin2007}. To correct for slit losses we
compared to the broad band photometry, and applied a constant scaling
to each of the three arms.

\section{Spectral Analysis}
\label{sec.spectralanalysis}
\begin{figure*}
\begin{center}
\includegraphics [scale=1]{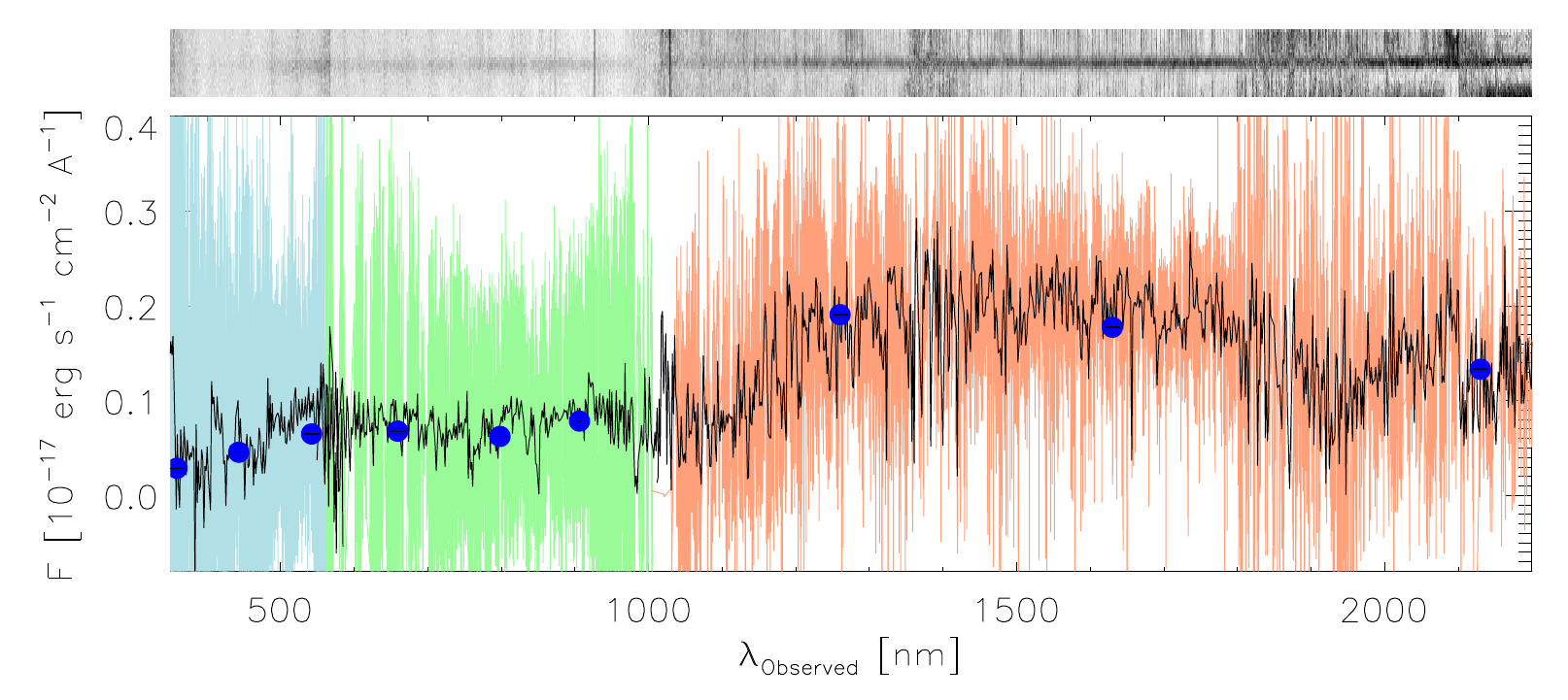}
\caption{Full 2D (top) and extracted 1D (bottom) X-shooter UVB (blue),
  VIS (green) and NIR (red) spectrum of UDS19627 (corrected for slit
  losses). The colored curves show the full resolution (0.5 {\AA})
  spectrum, the black curve is binned to a resolution of 10 {\AA},
  where the value in each bin is the weighted median of the individual
  pixel values. Also plotted are the broad band fluxes (blue
  circles). A number of absorption lines are detected (see also Figures
  \ref{fig.dust} and \ref{figure:finalfig}) \label{fig.threearms} }
\end{center}
\end{figure*}

In Figure \ref{fig.threearms} we show the full composite 3500--20000
{\AA} X-shooter spectrum, constructed by stitching together the flux
calibrated UVB, VIS and NIR arm spectra.  The spectrum confirms the
expectation from the broad band photometry, that the galaxy is a
massive quiescent galaxy at $z\sim2$, with a strong 4000 {\AA} break
in the J-band.  Interestingly the (flat) restframe UV continuum of the
galaxy is detected throughout the VIS and UVB arm, all the way down to
$\sim$ 3700 {\AA}. The overall shape of the continuum is characteristic of a
post starburst or E+A galaxy \citep[e.g.][]{quintero2004} rather than
of an early type galaxy. A number of strong absorption lines are
detected (Balmer series, Ca H+K, G-band). From the absorption lines we
derive a redshift of 2.0389$\pm$0.0004.

\section{Stellar Populations Properties}
\label{sec.stellarpops}
In order to constrain the galaxy stellar populations, we analyse the
X-shooter spectrum by means of a Monte Carlo library of 100000 model
spectra, based on
\citeauthor{BC03} (\citeyear{BC03}, hereafter BC03)
Simple Stellar
Population (SSP) models convolved with random star formation
histories. We adopt a \cite{Chabrier03} Initial Mass Function
(IMF). The SFHs are modeled with an exponentially declining function
on top of which random bursts of star formation can occur. Following
\cite{CF00}, attenuation by dust is modeled by means of the total
effective optical depth $\tau_V$ experienced by young ($<10^7$yr)
stars still in their birth cloud and the fraction $\mu$ of it
contributed by the ISM\footnote{We adopt the model library used is
  \cite{Salim05}, but we restrict the formation time to vary between
  the age of the Universe at the redshift of the galaxy (i.e. 3.2~Gyr)
  and 1.5~Gyr in order to reproduce the current quiescent nature 
  of the galaxy as witnessed by the lack of emission lines (we checked 
  that the precise choice of minimum formation time does not affect the 
  stellar mass estimates and only minimally the age estimates). 
  The other parameters defining the SFH and dust
  attenuation follow the same prior distributions as in
  \cite{Salim05}. The star formation timescale can vary between 0 and
  1~Gyr$^{-1}$. The probability of having a burst is set such that
  50\% of the models experience a burst in the last 2~Gyr. Bursts are
  parametrized in terms of the fraction of stellar mass produced,
  which is logarithmically distributed between 0.03 and 4, and their
  duration can vary between $3\times10^7$ and $3\times10^8$ yr. The
  metallicity of each model can vary between 10\% and 2 times solar
  and it is fixed along the SFH. The total effective optical depth
  $\tau_V$ can vary between 0 and 6, and $\mu$ can vary between 0.1
  and 1.}.

We adopt a Bayesian statistical approach in which all models 
in the library are compared to the observed galaxy. The likelihood of 
each model is computed as $\rm exp(-\chi^2/2)$ by comparing a selected 
set of observables with the model predictions. The posterior probability 
density function (PDF) of each physical parameter of
interest is computed by weighing all the models by their likelihood 
and marginalizing over all the other parameters 
\citep[see e.g.][]{Kauffmann03,Gallazzi05,Salim05}. The estimated parameter 
is taken from the median of the corresponding PDF, while the confidence 
interval is estimated from the $16^{th}$ and $84^{th}$ percentiles of the PDF.
In particular we derived PDFs of the luminosity- and mass-weighted ages, 
the stellar metallicity, the stellar mass and the parameters describing dust
attenuation.

In order to compare the data with the models we use a lower resolution version of the
X-shooter spectrum binned over 18 pixels \footnote{The binned spectrum
  is computed as the noise-weighted median flux in each bin and the
  error is obtained from the noise-weighted $16^{th}$ and $84^{th}$
  percentiles.} which corresponds to a rest-frame
$\Delta\lambda$=3{\AA} comparable to the spectral resolution of the
BC03 models in the optical range. We adopt a larger $\Delta\lambda$ of
16{\AA} in the rest-frame UV (UVB and VIS arms) to match the lower
resolution of the models in this regime. Finally, the observed
spectrum is corrected for foreground Galactic reddening ($A_V=0.0689$) 
adopting the
extinction maps of \cite{SFD98} and put in the rest-frame. The model
spectra are convolved with a range of velocity dispersions that
accounts for the measured velocity dispersion and associated
uncertainties (particularly relevant in the optical range and for
absorption index measures).

\begin{figure}
\centerline{\includegraphics[scale=0.5]{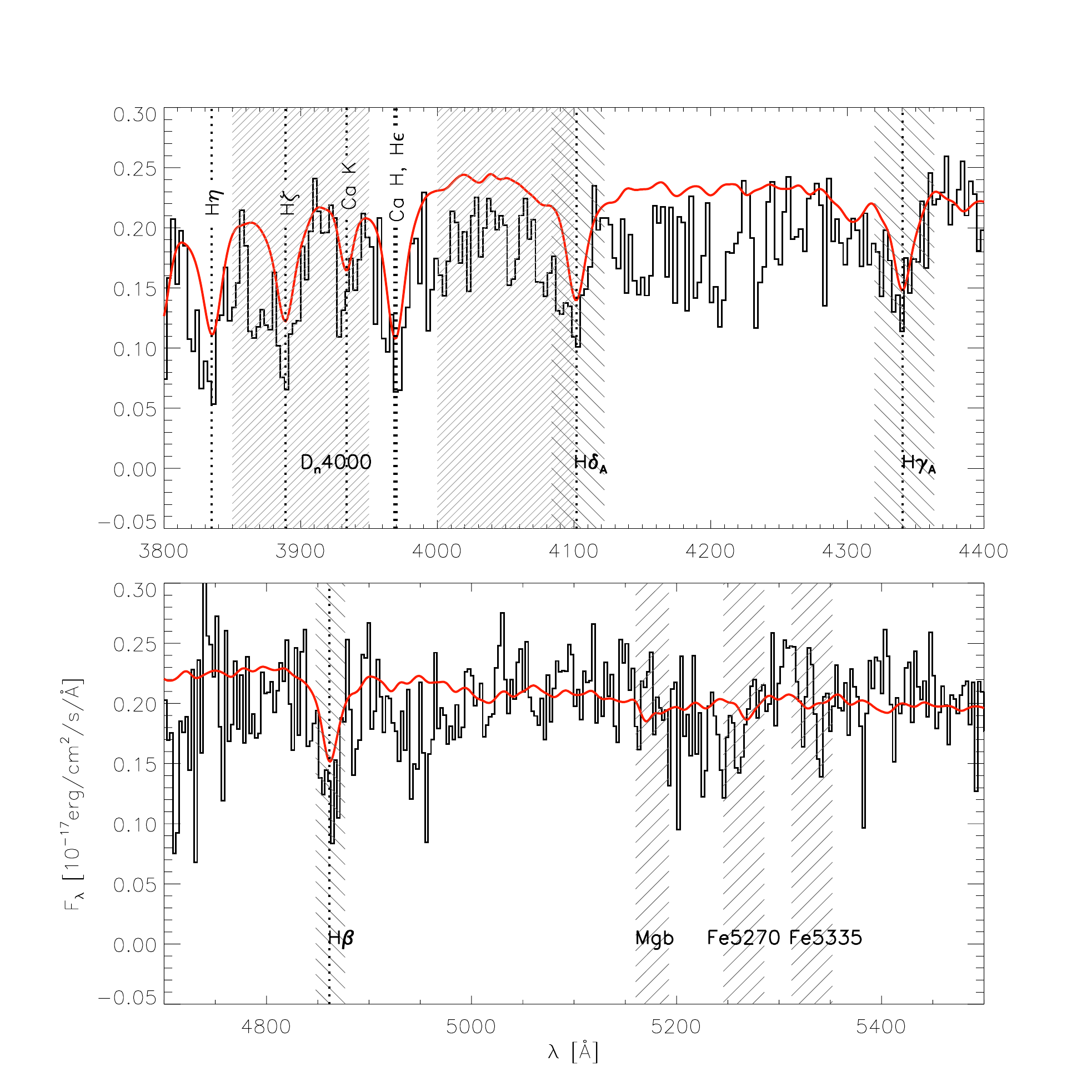}}
\caption{X-Shooter spectrum (binned to 3{\AA} per bin for
  illustration purposes) zoomed over the region of the absorption indices used in the
fit (the two bandpasses defining the 4000\AA-break and the central bandpass of the other indices are indicated by the hatched
regions). The vertical dotted lines indicate the main absorption features detected. The red spectrum is the bestfit model 
to the absorption indices, the UV flux ratio and the observer-frame J-H color.}
\label{fig.indices}
\end{figure}

In order to put constraints on the above mentioned parameters, we
combine information from individual stellar absorption features (see
Figure \ref{fig.indices}) with UV and optical colors. The indices we
analyze here are
chiefly sensitive to age and metallicity and minimally sensitive to
dust and $\alpha$/Fe abundance ratio \citep[this is important because our
models have solar-scaled abundances, e.g.][]{worthey1994,thomas2003} 
while the colors give constraints
on dust attenuation. Specifically, we interpret the stellar absorption
indices $D4000_n$, $H\delta_A+H\gamma_A$ and $[MgFe]^\prime$, the FUV
($1300-1780$\AA) to NUV ($1770-2730$\AA) flux ratio and the
observer-frame J-H color. The
parameter estimates and associated uncertainties (median and $16^{th}$
and $84^{th}$ percentiles of the PDF, respectively) are summarized in
Table~\ref{tab.paramcompare}.  We derive a luminosity-weighted
(mass-weighted) age of 0.8~Gyr (1.2~Gyr) with an accuracy of
$\lesssim$30\%. The rather weak and low-S/N Mg and Fe features do not
allow us to put strong constraints on stellar metallicity which is
estimated to be $\log(Z_\ast/Z_\odot)=0.02^{+0.2}_{-0.41}$. The
relatively large errors on stellar metallicity limit our ability to accurately
estimate dust attenuation, however the rather strong 4000\AA-break and
Balmer absorption lines in combination with the faint and flat UV
require a fair amount of dust in this galaxy. The attenuation by dust
is estimated to be $A_V=0.77^{+0.36}_{-0.32}$ in the optical and
$A_{FUV}=2.5^{+0.9}_{-0.8}$ in the UV. The attenuation in the optical
reflects the optical depth in the ISM, $\mu\tau_V$, experienced by
stars older than $10^7$ yr, while the attenuation in the UV traces
well the total effective optical depth $\tau_V$ of young stars.
Finally, the stellar mass, obtained by normalizing the models to the
observed luminosity between the rest-frame 5000--5800{\AA} (accounting
for dust attenuation), is estimated to be
$\log(M_\ast/M_\odot)=11.37^{+0.13}_{-0.10}$.

We tested how sensitive the derived parameters are to the
observational constraints adopted and found good agreement among the 
various age estimates and on the need for a significant amount of dust. We note 
though that the inclusion of UV constraints is particularly important for 
estimating dust attenuation. Our default fit combines the power of absorption 
features in alleviating parameter degeneracies and the sensitivity of UV and optical 
colors to dust attenuation. We 
take the standard deviation of the physical parameters obtained fitting different 
observational quantities as an estimate of the possible systematic uncertainties. 
The results are summarized in Table~\ref{tab.stelpop} while the comparison between 
different fitting methods is discussed in Appendix~\ref{appendix.stelpops}.

The estimated mass-weighted age of 1.2~Gyr implies that the galaxy
must have formed at $z>3$ ($z_f=3.2^{+0.5}_{-0.3}$ ), while the
luminosity-weighted age implies that star formation continued at least
for other 0.4~Gyr (i.e., until $z\sim2.7$). We look for the presence of
weak nebular emission lines after subtracting the bestfit stellar
continuum using the PLATEFIT code \citep{Tremonti04}\footnote{This
  code finds the bestfit non-negative linear combination of SSPs and
  the bestfit dust attenuation that describe the stellar continuum in
  regions free of emission lines. The bestfit stellar continuum (and
  any smoothed residuals) is then subtracted from the original
  spectrum in order to obtain the `pure' emission line spectrum.}. By
analyzing the residuals we find a 3$\sigma$ upper limit to the
H$\alpha$~ flux of $\rm 3.5\times10^{-17} erg/s/cm^2$. Adopting the
\cite{Kennicutt98} calibration between SFR and H$\alpha$ luminosity
($\rm SFR [M_\odot/yr] = 5.2\times 10^{-42} L(H\alpha) [erg/s]$ for
the Chabrier IMF), this translates into an upper limit to the current
SFR of $\rm 5.5 M_\odot/yr$. If we correct the H$\alpha$ luminosity
for the inferred dust attenuation in the optical, the SFR is $<10.3
\rm M_\odot/yr$ and the specific SFR is $<4.4\times10^{-11}yr^{-1}$.

Finally, we note that the derived stellar metallicity 
is consistent within the uncertainties with the metallicity of equally
massive $z=0.1$ galaxies \citep[see][]{Gallazzi05}. The
luminosity-weighted age is younger than the local population, as
expected, but, if we assume that the galaxy would evolve passively
since the redshift of observation until today, its present-day stellar
age would be fully consistent with the local age--mass relation.

\begin{deluxetable}{lcc}

\tablecaption{Physical parameter estimates and associated uncertainties 
obtained by fitting stellar absorption indices, the rest-frame UV and optical color. 
The parameter estimate is the median of the associated PDF, while the statistical 
uncertainty is measured from the $16^{th}$ and $84^{th}$ percentiles of the PDF. 
The last column gives the systematic uncertainty estimated by fitting different 
observational quantities (see Appendix~\ref{appendix.stelpops} for details).}\label{tab.stelpop}

\tablecolumns{3}
\tablewidth{0pc}
\tablehead{Parameter & Estimate & Systematic uncertainty}
\startdata

\noalign{\smallskip}
$\log(t_r/yr)$         & $8.90^{+0.10}_{-0.09}$   & 0.05 dex\\
\noalign{\smallskip}
$\log(t_m/yr)$         & $9.08^{+0.11}_{-0.10}$   & 0.03 dex\\
\noalign{\smallskip}
$\log(Z_\ast/Z_\odot)$ & $0.02^{+0.20}_{-0.41}$   & 0.06 dex\\
\noalign{\smallskip}
$\log(M_\ast/M_\odot)$ & $11.37^{+0.13}_{-0.10}$  & 0.12 dex\\
\noalign{\smallskip}
$\tau_V$               & $1.78^{+0.88}_{-0.63}$   & 0.33 \\
\noalign{\smallskip}
$\mu\tau_V$            & $0.64^{+0.32}_{-0.28}$   & 0.21 \\
\noalign{\smallskip}
$A_{FUV}$              & $2.52^{+0.89}_{-0.81}$   & 0.53 mag\\
\noalign{\smallskip}
$A_V$                  & $0.77^{+0.36}_{-0.32}$   & 0.27 mag\\
\noalign{\smallskip}

\enddata
\end{deluxetable}

\section{Surface brightness profile fits}
\label{sec.surfacebrightnessfits}
We fit the 2D surface brightness profile of the galaxy in the three
NIR UDS broad bands $J$, $H$ and $K$, using {\sc {galfit}}
\citep{peng2002}. We assumed a \cite{sersic1968} model and
allowed the Sersic parameter $n$, the effective radius $r_e$, the axis
ratio $ab$ and the position angle $PA$ to vary freely. The point
spread functions (PSFs) were estimated from high S/N stacks of stars
extracted from the UDS survey \citep[as described in][]{williams2010}.
In Figure \ref{fig.images} we show the K-band image of the galaxy and
illustrate the fitting process. There is a foreground
($z_{phot}=0.58$) galaxy $\sim3 \arcsec$ NW of UDS19627, which can
potentially influence the fit. We perform three different fits to
estimate the uncertainty it introduces. 1: Model the two galaxies
simultaneously, 2: mask out the foreground galaxy and fit the main
galaxy, 3: model and subtract the secondary galaxy (masking out the
main galaxy), and fit the main galaxy in cleaned image.  In Table
\ref{tab.galfit} we list the best fitting parameters of the main
galaxy for the three fits. The parameters are slightly sensitive to
how the secondary galaxy is taken into account in the fit, but agree
within the formal $3 \sigma$ uncertainties.

We adopt the best fitting parameters of Fit 1 in the remainder of the
paper, and include the standard deviation of the parameters in the three
fits in the error estimates, in addition to the formal fitting errors
from {\sc galfit}. These parameters are printed in bold face
in Table \ref{tab.galfit}.
In the same way we also derive  the best fitting parameters
and errors in the H and J bands. The Sersic parameters agree in the
three wavebands. The circularized sizes derived in the H and J band
are slightly larger, but more uncertain and agree within $2\sigma$.

\begin{deluxetable*}{l l l l l l l l}
  \tablecaption{Parameters of the 2D surface brightness
    distribution derived with {\sc galfit}.  $a_{\rm{e}}$ is the
    effective semi-major axis, n is the Sersic parameter,
    b/a is the axis ratio, $r_{e,c}=a_{\rm{e, major}}*\sqrt{b/a}$ is
    the circularized effective radius.  Fit 1-3 in the three first
    rows are done in the K-band, dealing with the foreground neighbor
    galaxy in different ways to estimate its effect on the derived
    parameters of the main galaxy. In Fit 1 the two galaxies are
    modelled simultaneously. In Fit 2 the main galaxy is fit with the
    neighbor galaxy masked out. In Fit 3 the main galaxy is modelled
    in an image where the neighboring galaxy has first been modelled
    and subtracted. The errors quoted for the three fits are those
    returned by {\em galfit}. The next three rows, labeled ``best
    fits'', are the parameters of the main galaxy from simultaneous
    fits in the K, H and J bands. The errors quoted here include the
    standard deviations from of the best fitting parameters from Fit
    1-3.  } \tablecolumns{7} \tablewidth{0pc} \tablehead{
    \colhead{Fit} & \colhead{Band} & \colhead{Mag} &
    \colhead{$a_{{\rm e}}$[$\arcsec$]} & \colhead{n} & \colhead{b/a} &
    \colhead{$\chi^2$} & \colhead{$r_{e,c}$ [kpc]} } \startdata 1 & K &
  $20.24 \pm 0.01$ & 0.44 $\pm 0.01$ & 1.47 $\pm 0.09$ & 0.74 $\pm
  0.02$ & 0.948 &
  2.77 $\pm 0.06$ \\
  2 & K & $20.18 \pm 0.01$ & 0.47 $\pm 0.01$ & 1.67 $\pm 0.07$ & 0.77
  $\pm 0.02$ & 0.963 &
  3.04 $\pm 0.07$\\
  3 & K & $20.26 \pm 0.01$ & 0.44 $\pm 0.01$ & 1.42 $\pm 0.07$ & 0.73
  $\pm 0.01$ & 1.073 &
  2.70 $\pm 0.05$\\
  \cutinhead{Bestfit with systematic errors}

{\bf1} & {\bf K} &   $\mathbf{20.24 \pm 0.04}$  & {\bf 0.44$\pm {\bf 0.02}$} & {\bf 1.47$\pm  {\bf 0.15}$} & {\bf
  0.73 $\pm {\bf 0.03}$} & {\bf 0.948} &
{\bf 2.77 $\pm {\bf 0.11}$} \\
1 & H &  20.54 $\pm0.11$  & 0.48$\pm 0.05$ & 1.43$\pm 0.35$  & 0.81 $\pm 0.03$ & 0.938 &
3.24 $\pm 0.37$\\
1 & J  &   20.99 $\pm0.07$  &0.50$\pm 0.03$ & 1.46$\pm0.23$   & 0.80 $\pm 0.05$ & 0.937 &
3.20 $\pm 0.20$\\
\enddata
\label{tab.galfit}
\end{deluxetable*}

\subsection{MIR emission}
The UDS was observed as part of the Spitzer UKIDSS Ultra Deep
Survey (PI Dunlop). At 24$\micron$, the position UDS19627 is blended
with a bright source coinciding with the nearby  $z=0.58$ galaxy (red cross in Figure
\ref{fig.images}). We model and subtract this source with galfit using
a bright nearby point source as PSF. The galaxy is not detected in the residual
image. In an $7.5\arcsec$ aperture we derive a flux of $0.1 \pm 20 \mu
Jy$, where we
have included the typical error in this size aperture quoted in the
Spuds survey documentation.\footnote{see
  http://irsa.ipac.caltech.edu/data/SPITZER/SpUDS/documentation/README.txt}. The
non detection rules out significant amounts of obscured star formation or
AGN activity. Using the method of described in \cite{franx2008} we derive a $2\sigma$ upper
limit on the star formation rate $SFR \lesssim 40 M_{\odot} yr^{-1}$.

\section{Stellar Velocity Dispersion}
\label{sec.velocitydispersion}
The presence of several strong stellar absorption features in our
spectrum allows for high quality measurement of the stellar velocity
dispersion. This determination is based on the assumption that the
observed line broadening is dominated by disordered motions of the
stars and that the contribution due to bulk flows or rotation is
minimal. We have therefore fit the spectrum using a set of
resolution-matched stellar template spectra to model the effect of the
velocity broadening.

\subsection{Penalized Pixel Fitting}

\begin{figure}
\begin{center}\includegraphics[scale=0.37,angle=0]{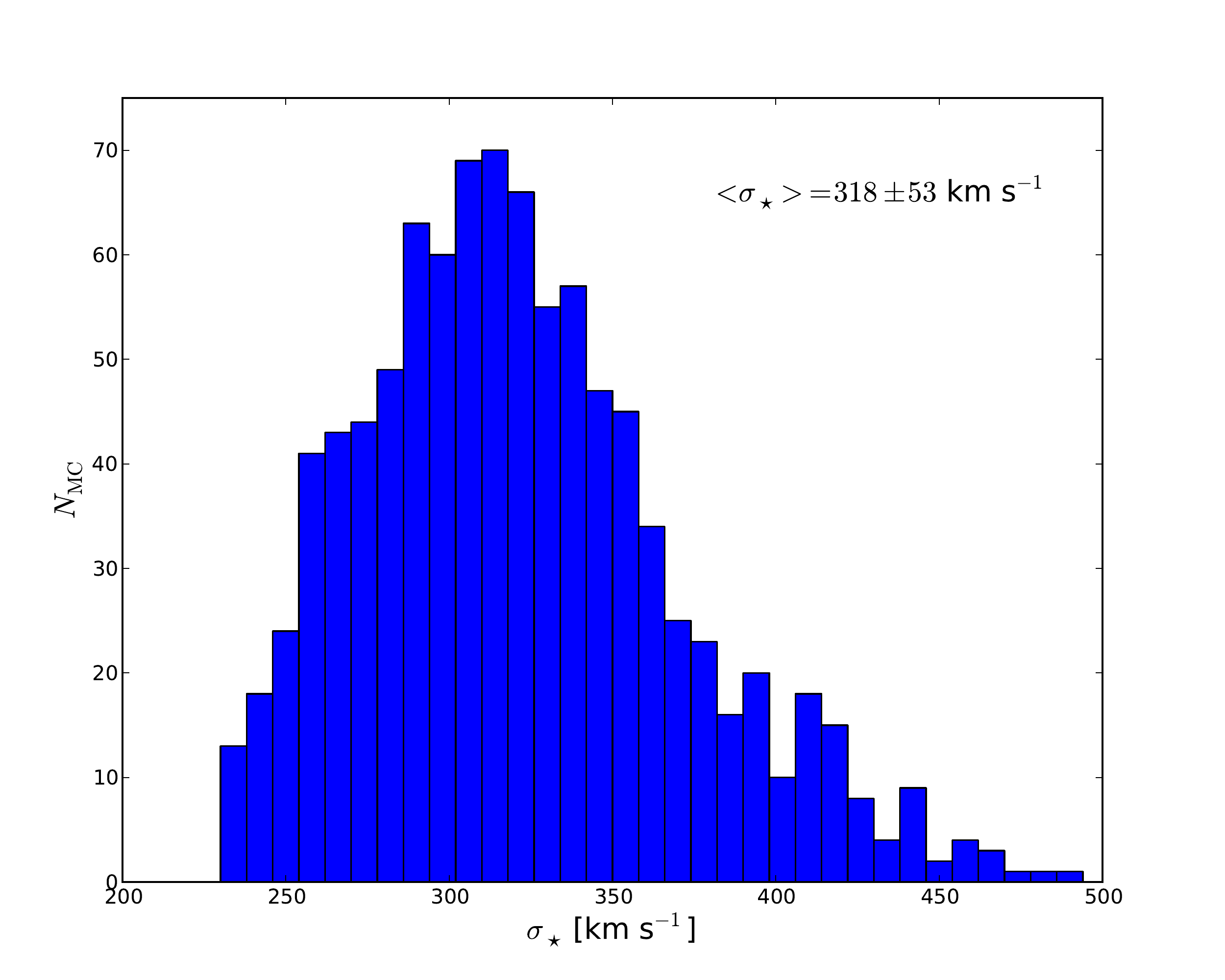}
\caption{Histogram of best-fit velocity dispersions for the 1000
  randomized realizations of the pPXF fitting. Each fit is produced by
a new linear combination of the initially chosen nine stellar
templates.}
\label{fig.ppxf}
\end{center}
\end{figure}

The public penalized pixel fitting code, {\sc{pPXF}}, developed by
\cite{cappellari2004}, has been used in the literature to fit velocity
dispersions of high redshift galaxies \citep[e.g.,][]{cappellari2009,
  onodera2010, vandesande2011}. We have run pPXF with the MILES
empirical stellar library \citep{sanchez-blazquez2006,
  falcon-barroso2011}.  As input, pPXF takes the raw X-Shooter
spectrum (excluding the observed $K$-band) along with all 985 stellar
spectra from MILES. We have shifted the observed spectrum to the
rest-frame, smoothed it with a Gaussian to match the instrumental
resolution of the templates and used logarithmic binning to put all
spectra on the same wavelength grid.  The code then constructs an
optimized linear combination of a small subset of the 985 spectra, in
this case, nine templates are chosen and combined. Using only this
subset we have run 1000 fits where the target spectrum is randomized
within the errors and the weights for the nine input templates are
allowed to vary (including zero weight). The resulting distribution of
stellar velocity dispersions are shown in Figure~\ref{fig.ppxf}. The
best fitting velocity dispersion from this analysis is
$\sigma_*=318\pm53$ km s$^{-1}$.

The inferred velocity dispersion changes in relation to the radii at
which it is measured. Using Equation 1 from Cappellari et al. (2006)
and assuming that the scaling from the spectrum to the broad-band
fluxes is only due to missing light from the profile, we infer the
correction to $\sigma$ at $r_e$ to be only a few percent. 

We performed an independent test of the derived velocity dispersion
through a  chi-squared analysis in which we fitted gaussian convolved, stellar
template spectra of five different types (AV, F5V, G8IV, K0III,
K0V) observed with X-shooters UVB arm, which corresponds approximately to the same restframe
wavelength range as sampled by the NIR arm for the z=2 galaxy. 
The AV and F5V stars and linear combinations hereof provide good fits
with velocity dispersions in the range $260-300$ km s$^{-1}$ with
typical uncertainty of $50 $km s$^{-1}$, in good agreement with the
results from pPXF. The analysis is described in detail in
Appendix \ref{appendix.vdisp}.

\subsection{Dynamical mass}
We estimate the dynamical mass of the galaxy from the derived velocity
dispersion and effective radius (in the $K$-band):
\begin{equation}
M_{dyn}=\beta r_e \sigma_e^2/G,
\label{eq.mdyn}
\end{equation}
where $\beta$ is a constant that depends mainly on the structure of
the galaxy.  A value of $\beta=5$ is commonly used in the literature
\citep[e.g.][]{jorgensen1996}.  From detailed analysis and modeling of
spatially resolved kinematic observations of a sample of 25 local E
and S0 galaxies and comparison to detailed modeling of stellar orbits,
\cite{cappellari2006} found that $\beta=5.0\pm0.1$ accurately
reproduces galaxy dynamical masses.  
Adopting this value, we find $M_{dyn}=
3.26 \pm 0.99\times10^{11} M_{\odot}$.  From simulations it is
predicted that $\beta$ will depend on the structure of the galaxy,
increasing with smaller Sersic n values \citep[][]{cappellari2006}.
For the best fitting $n=1.47$, a value $\beta\sim 7.5$ is predicted.
If we adopt this value we find $M_{dyn}= 4.88 \pm 1.49 \times10^{11}
M_{\odot}$, but as this dependency on structure has not been verified
observationally \citep{cappellari2006}, we will adopt $\beta=5$ in
this paper, which also makes comparison to the literature more straight
forward (but see Section \ref{sec:caveats}).

\subsection{Spectral Classification of UDS19627}
The spectrum of UDS19627 has the characteristic ``shark-tooth'' shape
of a post starburst galaxy, consistent with the low derived SFR, and
constraints on the stellar age and the prominent Balmer series
absorption lines.  The derived magnitudes (see Table \ref{tab.galfit})
corresponds to an $L_K\sim15L_K^*$ galaxy \citep[assuming
$K^*_{z=2}=21.3$, ][]{marchesini2007}.  The Sersic fit to the surface
brightness distribution yields a low $n=1.47$, also more similar to
what is found in late type (and post starburst) galaxies, than what is
found in local early types.

\section{Sample of $z\sim 2$ Galaxies With Measured Velocity Dispersions}
\label{sec.sample}
Prior to the results presented here, velocity dispersions of three
quiescent, massive $z\sim2$ galaxies have been published. We include
the properties of these galaxies in our analysis. Properties of their  stellar populations have all been derived assuming a Chabrier IMF. To minimize
systematics errors, we recalculate the dynamical masses 
using effective radii derived from restframe optical, and assuming the
same value for $\beta=5$.

\begin{deluxetable*}{l l l l l l l l l}
\tablecaption{Properties of the sample of $z\sim2$ quiescent galaxies with
  measured velocity dispersions.}
\tablecolumns{9}
\tablewidth{0pc}
\tablehead{
\colhead{ID} & \colhead{z} & \colhead{re} & \colhead{n} &
\colhead{$\log (M_{*}$)} &
\colhead{log(Age)} &\colhead{$\sigma_{obs}$} & \colhead{$\log(M_{dyn}$)} &\colhead{$\sigma_{inf}$}\\ 
\colhead{} & \colhead{} & \colhead{[kpc]} & \colhead{} & \colhead{[M$_{\odot}$]} &
\colhead{[Gyr]} &\colhead{[km s$^{-1}$]} & \colhead{[$M_{\odot}$]} &\colhead{[km s$^{-1}$]}
}
\startdata
UDS 19627 & 2.0389 &  2.77$^{+0.11}_{-0.11}$ & 1.47$^{+0.15}_{-0.15}$ & $11.37^{+0.13}_{-0.10}$
& 8.90$^{+0.10}_{-0.09}$ & 318$^{+53}_{-53}$ & 11.51 $^{+0.11}_{-0.11}$ & 339\\
\smallskip 
 NMBS-C7447 & 1.800 & 1.64$^{+0.15}_{-0.15}$ & 5.30$^{+0.40}_{-0.40}$ & $11.18^{+0.20}_{-0.20}$
& 8.60 & 294$^{+51}_{-51}$ & 11.23 $^{+0.11}_{-0.12}$ &  281\\
\smallskip 
COSMOS-254025 & 1.820 &  2.40$^{+0.40}_{-0.40}$ & 2.50$^{+0.40}_{-0.40}$ & $11.54^{+0.06}_{-0.06}$
& 9.18$^{+0.30}_{-0.18}$ & $<$326 & $<$11.47 &  354 \\
\smallskip 
MUSYC 1252-0 & 2.186 &  0.78$^{+0.17}_{-0.17}$ & 3.20$^{+0.90}_{-0.90}$ & 11.36$^{+0.04}_{-0.04}$
& 9.24$^{+0.10}_{-0.13}$ & 510$^{+165}_{-95}$ & 11.37$^{+0.25}_{-0.13}$ & 469\\
\enddata
\label{tab.sample}
\end{deluxetable*}

\subsection{COSMOS-254025}
\cite{onodera2010} reports $\sigma_* <$ 326 km s$^{-1}$ derived from a
Subaru/MOIRCS spectrum of a $z=1.82$ galaxy selected from the COSMOS
survey. This is an upper limit due to the limited resolution of the
spectrograph.  Using the results of surface brightness fits by
\cite{mancini2010} ($r_e=5.79\pm0.61$ kpc and $n=4.14$) from HST/ACS
F814W band observations, they derive a dynamical mass of $M< 7 \times
10^{11} M_{\odot}$.
For this analysis we measure the effective radius
in NIR images from the Ultravista survey (McCracken et al, in prep.),
sampling the restframe optical, which is where the velocity
dispersion is measured, and is a better tracer of the
stellar mass than the ACS I band (restframe UV),  as $z\sim2$ galaxies can have strong
  morphological k-corrections \citep[e.g.][]{toft2005, cameron2011}.
Indeed, comparison of the ACS and Ultravista images shows that the galaxy is extended and clumpy
in the restframe UV, but smoother and more centrally
concentrated in the restframe optical. 

Using an identical approach as described in Section \ref{sec.surfacebrightnessfits} we
run galfit on the deep J, H, and K band UltraVISTA survey observations
(McCracken et al, in prep.), masking out neighboring galaxies, and
using a bright nearby star as PSF model, we find a mean effective
radius of $r_e=2.4\pm 0.4$ kpc, and $n=2.5 \pm 0.4$. The derived $r_e$
is $\sim 60\%$ smaller than the value derived by \cite{mancini2010} in
the $F814W$ band. 
While the galaxy is only marginaly resolved in the UltraVISTA data,
sizes measured for galaxies using the procedure followed have been
demonstrated to be reliable for galaxies with intrinsic sizes several times smaller
than the PSF \citep{trujillo2006,toft2009} when the galaxy is
  detected at high S/N and a good PSF model is available, as is the
  case here. Note that the extend of the galaxy in the ACS image
  is twice the size of the Ultravista PSF (FWHM$\sim0.7\arcsec$), so if the restframe optical
  morphology had been similar it would have been well resolved in the
  Ultravista images. 
%The reason for the size difference may not solely
%be a wavelength effect, as we in the UltraVISTA Y band, which is the
%closest in central wavelength to the I814W band, derive a $r_e=3.0 \pm
%0.1 $kpc. 

 With our measured $r_e$ the upper limit on the dynamical
mass becomes M$_{dyn} < 2.97 \times 10^{11} M_{\odot}$.  From the
surface brightness fits to the UltraVISTA data we derive an H-band
magnitude of 18.42 $\pm 0.01$ and K = 19.81$\pm$0.01, corresponding to
an $L_K=22L_K^*$ galaxy \citep{marchesini2007}.  From a combined fit
of the spectrum and broad band photometry \cite{onodera2010} quote a
stellar mass of 3--4$\times10^{11}$M$_{\odot}$, and a luminosity
weighted age of 1-2 Gyr, corresponding to a formation redshift of
$z_f=2.5-4$.  The shape of the spectrum is typical of a post starburst
galaxy, with prominent Balmer absorption features.

\subsection{NMBS-C7447}
From an X-shooter spectrum of a redshift $z=1.800$ galaxy in the
COSMOS field, \cite{vandesande2011} derive a velocity dispersion of
$\sigma = 294 \pm 51 $km s$^{-1}$, a dynamical mass
M$_{dyn}=(1.7\pm0.5)\times10^{11}M_{\odot}$, and from HST/WFC3
observations in the F160W band, an effective radius of $r_e=1.64 \pm
0.15$ kpc and Sersic parameter $n=5.3\pm0.4$.  The X-shooter spectrum
has a typical post starburst shape, with prominent Balmer lines, and
the stellar population synthesis fits to the spectrum is consistent
with little ongoing star formation ($0.002M_{\odot} yr^{-1}$), a
relatively young age 0.4 Gyr, corresponding to a formation redshift of
$z_f\sim 2$ and a stellar mass $1.5\times10^{11}M_{\odot}$. In the
MUSYC catalog \citep{gawiser2006} it has $H=19.85$ and $K=19.75$,
corresponding to an $L_K\sim23L_K^*$ galaxy \citep{marchesini2007}.

\subsection{MUSYC 1252-0} 
\cite{vandokkum2009} report a $\sigma = 510^{+165}_{-95}km s^{-1}$
derived from a deep Gemini/GNIRS spectrum of a $z=2.186$ galaxy
selected from the MUSYC survey \citep{gawiser2006}.  It has $H=21.31$,
$K=21.03$, corresponding to a L=7L$_K^*$ galaxy
\citep{marchesini2007}.  Based on surface brightness fits to HST/NIC2
F160W observations, they find $r_e=0.78\pm0.17$kpc and $n= 3.2 \pm 0.9$
\citep{vandokkum2008}.  From these numbers, and assuming $\beta$=5, we
calculate $M_{dyn}=2.4^{+1.9}_{-0.8} \times 10^{11} M_{\odot}$ which
is identical to the value quoted in \cite{vandokkum2009}.  The shape
of the GNIRS spectrum is consistent with old, evolved, quiescent
stellar population, with little ongoing star formation (SFR$\sim 1-3
M_{\odot} yr^{-1} $) and age 1.3--2.0 Gyr, corresponding to a formation
redshift of $z_f \sim 4-7$ \citep{kriek2009}, quite different from the
younger, post starburst spectra of the other three $z\sim2$ galaxies
considered.  This galaxy also has a significantly larger velocity
dispersion and smaller size than the other three galaxies (see
Figure \ref{fig.sdss}).

\subsection{Comparison to Low Redshift Galaxies }
\label{sec.sdss}

\begin{figure*}
\begin{center}
\includegraphics[scale=1,angle=0]{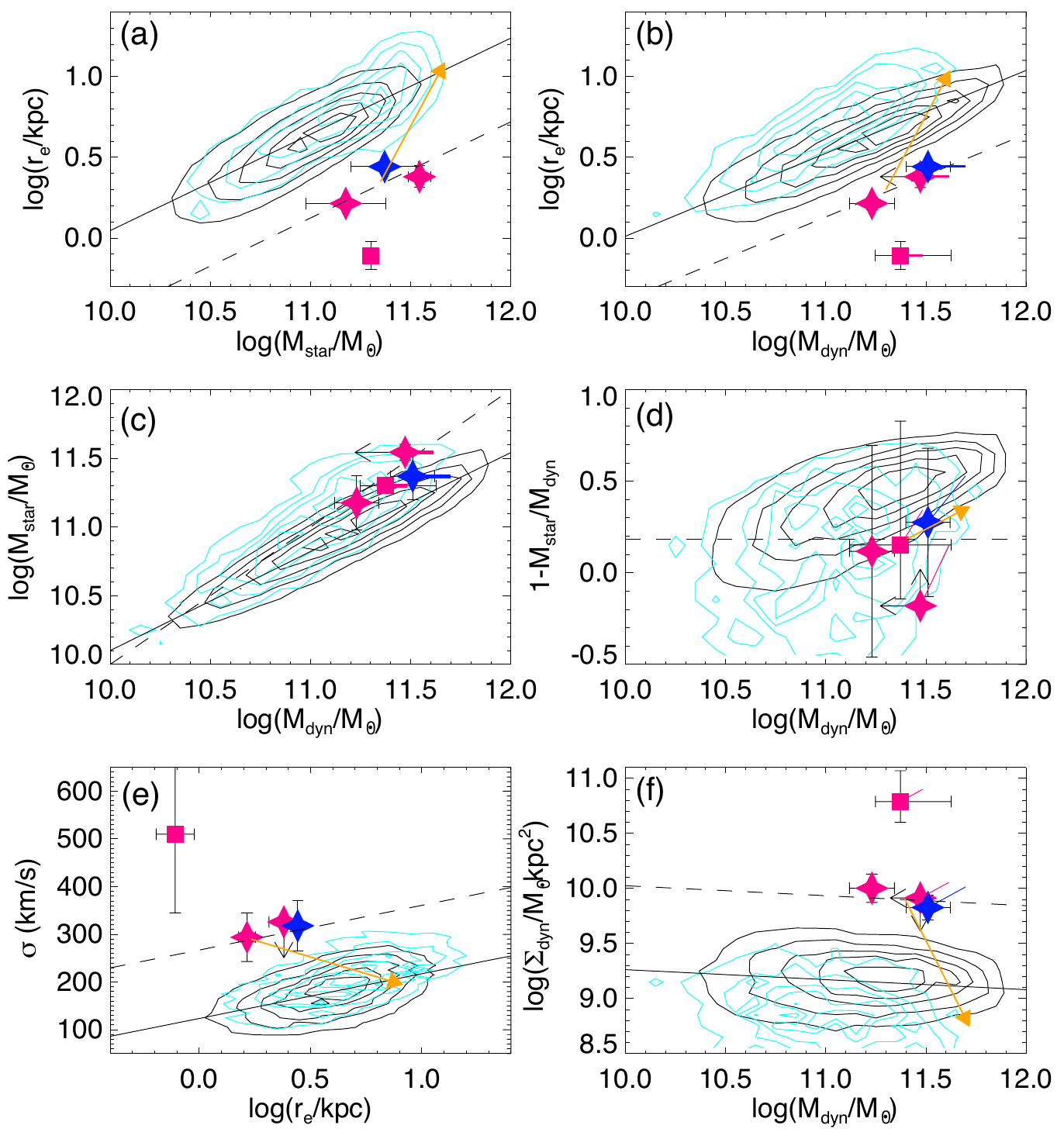}
\caption{Properties of the four $z\sim2$ massive quiescent galaxies
  with measured velocity dispersions, compared to early type and post
  starburst galaxies in the local universe \citep[drawn from the SDSS
  NYU-VAGC]{blanton2005}. The blue star is the galaxy considered in
  this paper. The two red stars and the red square are the post
  starburst and evolved SEEDs from the literature.  The black/cyan
  contours are low redshift early type/post starburst SDSS galaxies.
  The full black lines are linear fits to the local galaxies, the dashed
  lines are linear fits to the high redshift points, with the slope
  fixed to the local values. The orange arrows represent the evolution
  due to merging predicted derived from the cosmological simulations
  of \cite{oser2012}. The colored lines show the effect on the results
  of using the expression for $\beta(n)$ of \cite{cappellari2006} when
  calculating the dynamical masses rather than assuming $\beta=5$ (see
  Section \ref{sec:caveats}).
 {\bf Top left:} Stellar mass vs effective radius
  {\bf Top right}: Dynamical) mass vs effective radius. {\bf Middle
    left:} Dynamical mass vs stellar mass. {\bf Middle right} Central
  dark matter fraction ($1-M_{*}/M_{dyn}$) vs dynamical mass. {\bf
    Bottom left:} velocity dispersion vs effective radius, {\bf Bottom
    right:} Surface (dynamical) mass density vs dynamical mass. }
\label{fig.sdss}
\end{center}
\end{figure*}

In Figure \ref{fig.sdss} we compare the properties of the four $z \sim
2$ galaxies with measured stellar velocity dispersions (colored
points) to local $z<0.2$ early type (black contours), and post
starburst (cyan contours) galaxies from the SDSS New York University
Value-Added Galaxy Catalog \citep[NYU-VAGC][]{blanton2005}. The former
galaxies are selected using the criteria: SFR$<1$ M$_{\odot}yr^{-1}$,
and Sersic $n>3$, the latter using the criteria of \cite{goto2005}.
Effective radii are measured in the $i$-band.

Three of the $z\sim2$ galaxies (UDS19627, NMBS-C7447 and
COSMOS-254025, represented by stars) follow a mass-size relation with
similar slope as for local early type galaxies, but offset to smaller
sizes at a given mass (in the following we will refer to these three
galaxies as the post starburst SEEDs). One galaxy, MUSYC 1252-0
(square), is an outlier with respect to these (we refer to this galaxy
as the evolved SEED).

In Figure~\ref{fig.sdss} (a) we show the relation between $r_e$ and
stellar mass. The $z\sim2$ post starburst galaxies follow a relation
similar to the local early type relation (solid line), but offset by a
factor of $3.3\pm0.2$ to smaller $r_e$ for a given stellar mass
(dashed line). This offset is similar to what has previously been
found for larger samples of stellar mass selected quiescent galaxies
at $z\sim2$, with photometric masses and redshifts,
\citep[e.g.][]{toft2009}, but smaller than the factor of five offset
found for color selected $z\sim2$ quiescent galaxies
\citep[e.g.][]{toft2007, zirm2007, cimatti2008, vandokkum2008},
exemplified in these plots by the color selected MUSYC 1252-0 galaxy.
The stellar mass-size relation of low-$z$ post starburst galaxies is
on average shifted along the early-type relation to larger masses and
sizes.  In Figure~\ref{fig.sdss} (b) we show the same relation for
dynamical mass versus effective radius.  The $z=2$ post starburst
galaxies follow a relation similar to the stellar mass-size relation, but
with a smaller offset and scatter.  $r_e$ for the $z\sim2$ post
starburst galaxies are on average $2.5\pm0.2$ times smaller than
local early type galaxies of similar dynamical mass.  As illustrated
in Fig \ref{fig.sdss} (c) the difference in offset in (a) and (b) is
due to a difference in the derived stellar to total mass ratio.
%% describe reasons these could be different, concluding that DM
%% fraction is the best explanation? what about observational offsets?

For local early type galaxies there is a tight correlation between
$M_{dyn}$ and $M_{*}$, with the ratio of dynamical to stellar mass
(within $r_e$) increasing from the least massive to the most massive
galaxies \citep[see also][]{padmanabhan2004,gallazzi2006,
  taylor2010}. Under the assumption of structural homology, this can be
interpreted as an increasing dark matter fraction with mass.  The
local galaxies with ${\rm log}(M_{dyn}/M_{\odot}) \sim 10.5$ are thus
baryon dominated, while the most massive ${\rm log}(M_{dyn}/M_{\odot})
> 11.5$ have much larger dark matter fractions.  A similar trend has
been shown to be valid in early type galaxies out to $z\sim1$
\citep[e.g.][]{ferreras2005, rettura2006}.  While the $z\sim2$
galaxies are consistent with the local relation within the error bars,
it is striking that the best fitting values all fall close to the dashed
line ($M_{*}=M_{dyn}$) indicating a lower dark matter fraction than in
similar mass early type galaxies at lower redshift. We explore this
further in Figure\ref{fig.sdss} (d) where we plot $1-M_{*}/M_{dyn}$: a
proxy of dark matter fraction as a function of dynamical
mass (assuming homology).  The central dark matter fraction of the
$z\sim2$ galaxies scatter around a mean value $1-M_{*}/M_{dyn}=0.18\pm
0.20$, while the low redshift galaxies in the same dynamical mass
range have a mean of $0.46 \pm 0.23$.  Note from (b) that the
dynamical mass-size relation of the local post starburst galaxies is
shifted from the early type relation to larger sizes at a given
mass. 
From plot (c) and (d) it can be seen that the low dark-matter fraction
found in the $z\sim2$ galaxies is consistent with what is found in
local post starburst galaxies, which show a large range in
$M_{*}/M_{dyn}$, but on average have $M_{dyn}\sim M_{*}$.  The
observed low central dark matter fraction of the $z\sim2$ galaxies
compared to that in local early type galaxies could have implications for
the evolutionary path to lower redshift. We explore this further in
the discussion (Section \ref{sec.discussion}).

In Figure \ref{fig.sdss} (e) we plot the effective radius versus the
velocity dispersion. The average velocity dispersion of the $z\sim2$
post starburst galaxies is $1.8   \pm 0.5$ times larger than in low
redshift galaxies of similar size, while the velocity dispersion of
MUSYC 1252-0 is $4.1\pm2.1$ times larger than
local galaxies of similar size. The local post starburst galaxies have
a similar distribution as local early types but shifted to larger
sizes and velocity dispersions.

In Figure \ref{fig.sdss} (f) we plot the inferred (dynamical) mass
density within the effective radius $\Sigma=0.5 M_{dyn}/(\pi r_e^2)$,
versus the dynamical mass $M_{dyn}$. The $z\sim2$ post starburst
galaxies have average surface mass densities $5.8 \pm 1.2$ times
higher than the mean in local early types of similar mass. The
relation for the local post starburst galaxies are shifted to smaller
surface mass densities. As in the dynamical mass-size plot, the offset
of the z=2 post starburst galaxies from the local post starburst
relation is similar to the offset of the evolved z=2 galaxy from the
local early type relation.

\subsection{Evolution Through Merging?}

Dry merging has been suggested as a mechanism to ``puff up'' the SEEDs
and evolve them into agreement with the mass-size relation observed at
low redshift \citep{naab2007,naab2009,bezanson2009,newman2011}.
Previous studies of the stellar mass-size relation have not been able
to unambiguously distinguish between different merging scenarios, but
minor merging is the most promising process. With the
added dynamical information we can now start to test different merger
scenarios in greater detail than what is possible with just stellar
masses and sizes.  The low inferred dark matter fraction of the
$z\sim2$ galaxies, compared to early types at low redshift, supports
the merging scenario, as merging can redistribute the dark matter
within the effective radius \citep[][]{boylan-kolchin2005, oser2012}.
Furthermore, if the dark matter profile is more radially extended than
the stars, increasing $r_e$ by adding stars in the outskirts can lead
to an increased measured dark matter fraction by up to a factor of 2-3
because areas with higher dark matter content is included within the
effective radius (Hiltz et al, in prep., T. Naab, private
communication).

Based on simple virial arguments \cite{bezanson2009} and
\cite{naab2009} show that dry merging can lead to an increase in size
of the remnant, and that minor merging is likely to be the dominant
process, as this is the most effecient process for size growth, in
terms of added mass needed for the observed size evolution (about a
factor of 2). Cosmological simulations studying the effect of merging
on the size evolution of massive galaxies, show that dominant mode of mass growth
is minor merging, with a typical mass weighted ratio of 1:5
\citep{oser2012}. The average integrated size growth between z=2 and 0
for $M\sim10^{11}M_{\odot}$ galaxies in these simulations is a factor
of 5-6, and an average mass growth of a factor of 2.1. The velocity
dispersion in the remnants are approximately 30\% lower than in their
progenitors at z=2. This evolution is almost identical to what is
predicted from the  simple virial approximations for a similar mass
growth. Interestingly, in these simulations galaxies at z=2 typically
have low dark matter fractions at $z=2$ (0.1-0.3, similar to what is
derived here for the observed $z\sim2$ galaxies), and these
increase with time as a consequence of the merging by approximately a
factor of 2.  In Figure \ref{fig.sdss} we
plot with arrows the evolution from the cosmological
simulations of \cite{oser2012}. For the $z\sim2$ post starburst
galaxies, these predictions provide a good match to the observed
evolution of the stellar mass and velocity-size relations. However,
they fail to reproduce well the evolution in the dynamical mass-size
and dynamical mass density - mass relations where the galaxies end up
above and below the local relations, respectively, as a consequence
of the evolution of dark matter fraction in the simulations not being sufficient
to account for the observed evolution.
For the evolved $z\sim2$ galaxy additional structural evolution is needed.
Some of the additional evolution needed in the dark matter fraction
may be attributed to merging with 
galaxies with higher dark matter fractions than those included in the simulations.  In the local
universe dwarf galaxies can be heavily dark matter dominated, with
stellar masses $\sim10^7M_{\odot}$, and dynamical 
masses $\sim 10^8-10^9M_{\odot}$, corresponding to dark matter
fractions as high as $95\%$ \citep[e.g.][]{carignan1989,persic1996,
  strigari2010}.  If the $z\sim2$ galaxies grow via minor merging
with primarily dark matter dominated dwarfs, it could be an efficient
way of increasing the dark matter fraction in the remnant.

\subsection{The Fundamental plane at $z\sim2$}
\label{sec.fp}

\begin{figure}
\begin{center}
\includegraphics[scale=0.63,angle=0]{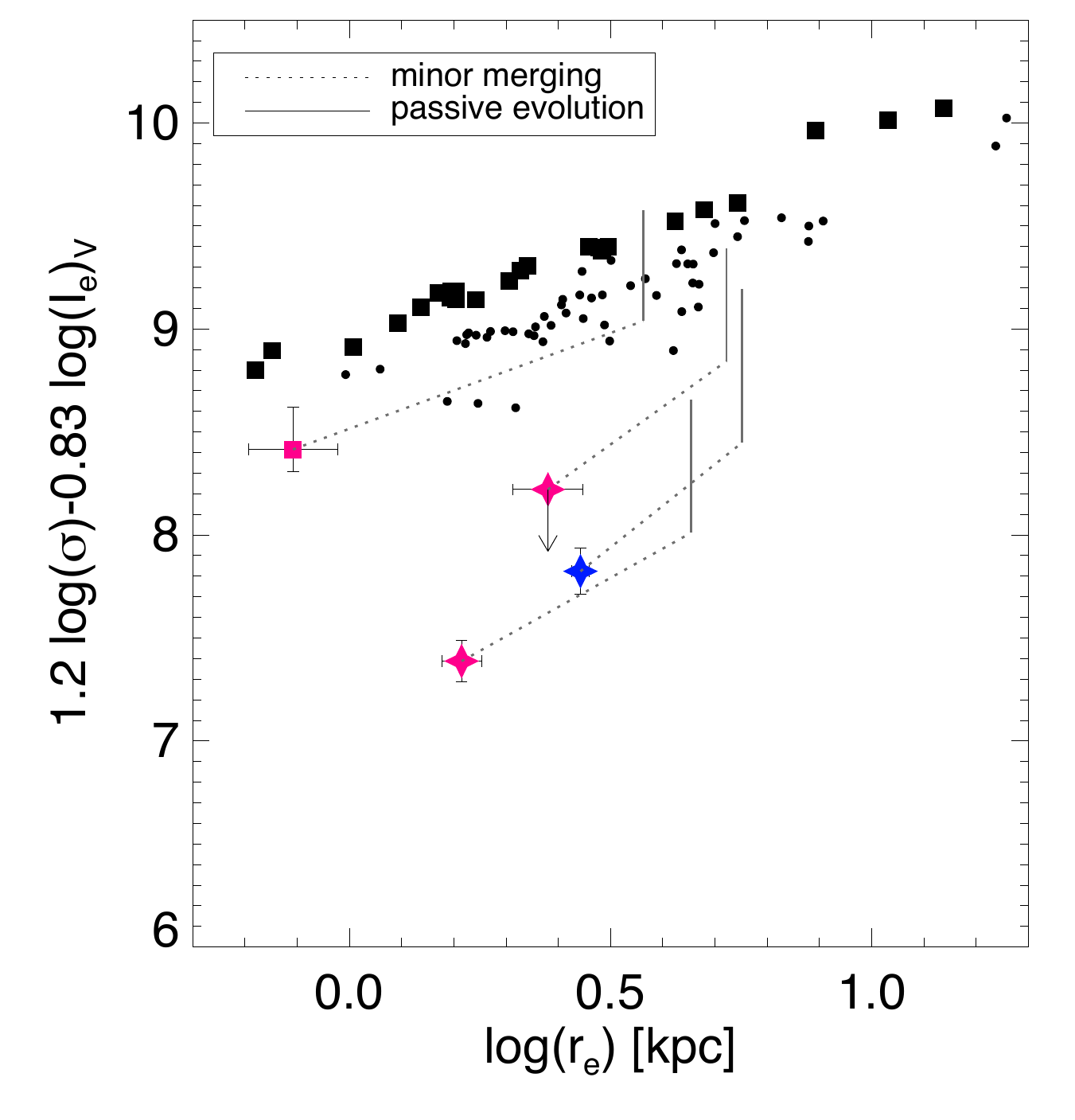}
\caption{Comparison of the restframe V-band fundamental plane of the
  $z\sim2$ galaxies compared to in the 
  z=0.02 Coma cluster \citep[black squares,][]{lucey1991}, and in the
  z=0.33 cluster CL1358+62  \citep[black circles,][]{kelson2000}. The stars are the post starburst SEEDs, the blue is
    the one considered in this paper, the square is the evolved SEED.
The $z=2$ galaxies are offset significantly from the
    local relation.  The  lines indicate the effect of luminosity and
    structural evolution considered between z=2 and 0. The dotted and
    dashed lines are the evolution predicted by the minor and major
    merging scenarios in Section \ref{sec.sdss}, and the full line is
  the effect of passive evolution of the M/L predicted by the best
  fitting stellar population synthesis models between their observed
  redshift, and today, assuming the best fitting formation redshift
  $z_f$, a Chabrier IMF and solar metallicity. These combined
  evolutionary effect are able to bring the evolved SEED close the local
  fundamental plane, but cannot fully account of the offset of the
  post starburst SEEDs}
\label{fig.fp}
\end{center}
\end{figure}

From the measured velocity dispersions, effective radii, and the mean
restframe surface brightness within the effective radius 
\begin{equation}
\left< \mu_e \right > = V_z-5\log(r_e)+2.5 \log(2\pi) - 2.5 \log( (1+z)^4) - A_V,
\end{equation}

we can for the first time probe the fundamental plane of  $z\sim2$
quiescent galaxies: 

\begin{equation}
\log(r_e) = \alpha \log(\sigma) - \beta \log(I_e),
\end{equation}
where  $V_z$ is the (AB) magnitude in the redshifted $V$ band, which following \cite{vandokkum1996} can be
estimated simply as $V_z=H+2.5$log$(1+z)$ (since the observed $H$-band resembles
closely the $V$-band redshifted to $z\sim2$), 
$r_e$ is the ($H$-band) effective radius, $\sigma$ is the velocity dispersion
in km s$^{-1}$ and $I_e=10^{(-0.4*\mu_e)}$. 

 Here we adopt  $\alpha=1.2$, and $\beta=0.83$, as derived by \cite{jorgensen1996}
  for a sample of 225 galaxies in 10 nearby clusters. Note that this
  quite strong  assumption for the 3D orientation of the fundamental
  essentially builds in homologous evolution in the following
  analysis, but its the best we can do given the paucity of $z\sim2$
  data points. 

In Figure \ref{fig.fp} we compare the rest-frame V-band fundamental
plane of the $z\sim2$ galaxies to the fundamental plane observed
locally in the $z\sim0.02$ Coma cluster \citep{lucey1991} and in the
$z\sim0.33$ cluster CL1358+62 \citep{kelson2000}. The galaxies have
been corrected for cosmological $(1+z)^4$ surface brightness dimming.
The slope of the $z=0.33$ fundamental plane is similar to in the Coma
cluster, but slightly offset.

While it is not possible to derive robust estimates of slope or
scatter of the $z\sim2$ galaxies, we note that they show a
significantly larger scatter than the local relation, and a large
offset from it.  As in Figure \ref{fig.sdss}, the evolved and the post
starburst $z\sim2$ galaxies fall in different parts of the plot. Assuming the
local slope, the post star burst galaxies are on average offset by
$1.6 \pm 0.2$ dex from the local relation.  The evolved SEED is closer to
the local relation.

As the velocity dispersion traces the total mass, and the surface
brightness traces the light, the offset between the fundamental planes
at different redshifts is often interpreted as an offset in
mass-to-light ratio caused by passive aging of the stellar
populations, which shifts the plane up, as the surface brightness
fades, while the velocity dispersion stays the same
\citep[e.g.][]{vandokkum2007}.

The offset observed for the $z=2$ post starburst galaxies is too large to be
explained by passive evolution. The solid lines represents the maximum
evolution of the surface brightness that can be attributed to passive
aging of the stellar population between z=2 and z=0, assuming solar
metallicity, a Chabrier IMF, and that the stars were formed in a
single stellar burst at their derived $z_f$. We note that this
relies on the ages being correct. If these are systematically
overestimated there could be room for more passive evolution.  It is
clear that in order to end up on the local relation the galaxies also
have to go through structural evolution.  The minor merging scenario
described above would also have an effect on the fundamental plane as
it changes the velocity dispersion and size, and thus also the surface
brightness of the galaxies. The expected shift of the galaxies due to
this process is indicated with dotted lines.

These evolutionary effects in concert brings the z=2 post starburst galaxies closer
to the local relation but still systematically under. The offsets
range from 0.2-1 Dex (with a mean of $0.5 \pm 0.4$). Interestingly, the evolved SEED ends up on the local relation through these processes, however
from Figure \ref{fig.sdss} we know that it needs to go through
additional structural evolution to end up on the local mass-size
relation.

A possible origin of the additional offset from the local fundamental
plane that can not be accounted for by passive evolution and minor
merging is that the evolution between z=2 and 0 may be non
homologous. This is not unlikely as we know that at least some of the $z=2$
galaxies have to change their surface brightness profiles from
exponential disk to deVaucouleurs profiles.  Local post starburst
galaxies have in some cases been shown to be offset by a similar
amount from the fundamental plane of local elliptical galaxies
\citep{yang2004}, which is attributed to the very young stellar ages
of those particular galaxies. At the older observed ages of the
$z\sim2$ galaxies considered here, this is however not a very likely
explanation.  Metallicity and/or age gradients in 
the $z\sim2$ galaxies could  be a potential source of uncertainty,
however higher resolution multicolor imaging is needed to adress this.

\subsection{Clues to the formation mechanism of SEEDs}
The observations presented here hold important clues to the formation
scenario of SEEDs which is still not well understood.  
The uniformly
old and extremely compact stellar populations suggest that the
majority of the stars formed in a major nuclear starburst at high
redshift. Simulations indicate that highly dissipational processes on
short timescales are plausible mechanisms for creating the compact
stellar populations \citep[e.g.][]{naab2007, naab2009}.  A possible
scenario is major gas rich mergers at high redshift
\citep[][]{wuyts2010}, in which the gas is driven to the center,
igniting a massive nuclear starburst ($>1000 M_{\odot} yr^{-1}$),
followed by an AGN/QSO phase which quenches the star formation, and
leaves behind a compact remnant \citep{hopkins2006}.  Sub-mm galaxies
(SMGs)/Ultra lumnious Infrared galaxies (ULIRGS) may be examples of
this process in action.  Observations show that most of them are major
mergers, with large amounts of dust enshrouded star formation, and
high central concentrations of molecular gas, comparable to the
density of stars in SEEDs
\citep[][]{greve2005,tacconi2006,tacconi2008, michalowski2011}.  The
compact structure, post starburst nature and (at least in the case of
the galaxy studied here), relatively high extinction, make them very
likely descendants of $z>3$ dusty SMGs. 
With the accurate ages obtainable from NIR spectroscopy of quiescent
$z\sim2$ galaxies, we can begin to constrain the number densities of
their progenitors as a function of redshift, e.g. if one assumes a
direct evolutionary link between quiescent
$z\sim2$ galaxies and $z>4$ SMGs their relative number densities can be
used to constrain the duty cycle of starformation to be $\sim 50$ Myr \citep[][]{capak2008}

\subsection{Caveats}
\label{sec:caveats}
As the $z\sim2$ galaxies are spatially unresolved in the spectroscopic
observations, it is not possible to determine if some of the
broadening of the absorption lines, which we interpret as velocity
dispersion, could be due to rotation. This may be a likely
scenario given that some of their surface brightness profiles are best fitted
by exponential disk like profiles, and there is evidence from recent
high resolution observations that a significant fraction of SEEDs may
have flattened disk like morphologies \citep{vanderwel2011}.  This
could lead to a systematic bias in their velocity dispersions, and
dynamical masses. This, in turn, would mean that the derived offset
from the local dynamical mass-size relation could be biased. Other
sources of possible uncertainties in the dynamical masses are that the
assumptions of homology and isotropy, needed to calculate the
dynamical mass from the velocity dispersion may not be fully valid.
Some studies have suggested that the sizes of compact $z\sim2$
galaxies could be underestimated due to faint profile
wings not detected due to cosmological surface brightness dimming
\citep[e.g.][]{mancini2010}, which would lead to the dynamical masses
being under estimated. However, considering the bright magnitudes of
the galaxies considered in this paper, and that ultra
deep studies tracing the surface brightness profile of compact
$z\sim2$ galaxies out to $>10r_e$ have failed to detect such wings
\citep[e.g.][]{szomoru2010, szomoru2011}, this is
 not likely to be a very significant effect compared the other potential systematic errors.   

The derived low dark matter fraction is sensitive to the
$\beta=5$ assumption made when calculating the
dynamical masses. From theory it is predicted that $\beta$ should depend on the
structure of the galaxies in a way that can be approximated through
the following dependency on the sersic $n$ parameter: $\beta =
8.87 -0.831n+0.024n^2$ \citep{cappellari2006}. The colored lines in Figure
\ref{fig.sdss} shows the effect of assuming this $\beta(n)$ correlation.
 The dynamical masses increase 3-50\% for the $z\sim2$ galaxies,
resulting in a higher mean average dark matter fractions ($0.29\pm0.18$).
However,  \cite{cappellari2006} showed that
there is no observational support for this correlation and real
galaxies (with n=2-10) are best fitted by $\beta\sim5$, we therefore
adopt this value.

Another potential uncertainty is the Initial Mass Function (IMF)
assumed in the stellar population synthesis modeling.  Theoretical and
observational studies have argued that the IMF may depend on redshift
and/or environment \citep[e.g.][]{blain1999, tumlinson2007,
  vandokkum2008b, wilkins2008, treu2010}.  If that is the case, 
assuming the same IMF for the $z=2$ and $z=0$ galaxies may introduce a
bias in the estimated stellar masses. 
Observations of $z < 1$ massive early type galaxies acting as
 gravitational lenses suggest that a Salpeter IMF provides a better fit
  than a Chabrier IMF \citep[e.g.][]{grillo2009, auger2010, sonnenfeld2011}.
Assuming a Salpeter IMF rather
than an Chabrier IMF for the $z\sim2$ galaxies would lead to
approximately 1.75 times higher
stellar masses \citep[e.g.][]{gallazzi2008}, resulting in stellar masses
significantly higher than the dynamical masses, strongly disfavoring a
Salpeter IMF for the $z\sim2$ galaxies.  The derived stellar masses of
the $z\sim2$ galaxies has a relatively weak dependance on whether or
not the IMF is bottom light, at the observed stellar ages \citep[$\sim1-2$
Gyr,][]{marchesini2009}, but at the older ages of local galaxies
($\sim 10$ Gyr) the derived stellar mass is more sensitive to the
underlying IMF, leading to a potential bias between the $z=2$ and
$z=0$ mass size relations.  As a consistency check, we calculate the
so called ``inferred'' velocity dispersion, from the stellar mass and
effective radius: $\sigma_{inf}^2=G M_{*}/\beta r_e$.  The derived
$\sigma_{inf}$ of the $z\sim2$ agrees well (within $\sim10\%$) with
the measured velocity dispersions (as expected since $M_{dyn}\sim
M_{*}$, see Table \ref{tab.sample}).

As evident from Figure \ref{fig.images}, UDS19627 is located close to a
foreground galaxy (angular separation $\theta=2.83\arcsec$). This galaxy is responsible for a gravitational lensing effect
that causes a bias in the derived brightness, derived stellar mass, size, and
dynamical mass of the $z\sim2$ background galaxy. To estimate this effect we construct a simple lensing model of the system. We assume for the foreground galaxy a photometric redshift $z_{l}=0.58$ and stellar mass $\log(M_{*,l}/M_{\odot})=10.32^{+0.07}_{-0.05}$, where both values are derived from broadband B, r ,i ,z, J and K band photometry, using the same model library as
for our main target. We then compare this stellar mass value to those of the SLACS lens galaxies \citep{bolton2006, grillo2010}
 to estimate the effective velocity dispersion $\sigma_{SIS}$ of a singular isothermal sphere (SIS) model that we adopt here to describe the total mass distribution of the lens. We find that a value of 145 km s$^{-1}$ would be typical for lens galaxies with stellar mass values similar to that of our foreground galaxy. For a SIS model, we recall that the Einstein radius $\theta_{Ein}$ is equal to 
\begin{equation}
\theta_{Ein} = 4 \pi (\sigma_{SIS}/c)^2 \,D_{ls}/D_{os}, 
\label{eq.einstein_radius}
\end{equation}
where $c$ is the speed of light and $D_{ls}$ and $D_{os}$ are the angular diameter distances between the lens and the source and the observer and the source, respectively. The magnification factor $\mu$ at an angular distance $\theta$, that is larger than $\theta_{Ein}$, from the center of the lens is given by 
\begin{equation}
\mu(\theta)=1 + \theta_{Ein} / (\theta - \theta_{Ein}).
\label{eq.magnification}
\end{equation}
Following the previous two equations, we
calculate a magnification factor of 10-20\% for the $z \sim 2$ galaxy. The
brightness and stellar mass are thus over estimated by this
amount and the size by the square root of
this. By applying this correction, we would get a 0.07 dex lower stellar mass and a 0.03 dex lower dynamical mass. Given the relatively small weight of the
lensing effect compared to the other uncertainties in our analysis and the approximations adopted in the lensing modeling, we decide not to correct our
results for the lensing effect. We note that lensing is likely
to be a general issue that needs to be taken into account when
analyzing samples of the very brightest $z \sim 2$ quiescent field
galaxies.

\section{Summary and Discussion}
\label{sec.discussion}

Since their initial discovery \citep{toft2005,daddi2005, trujillo2006,
  toft2007, zirm2007} the extreme properties of SEEDs have been
heavily debated, as galaxies with such properties were not expected
from galaxy evolution models, and they do not exist in the local universe
\citep{trujillo2009, taylor2010, shih2011}.  Initial worries that
their large inferred mass densities, could be caused by observational
biases and/or systematic uncertainties in their derived sizes have
since been ruled out through larger, better defined samples with
increasingly deeper, higher resolution imaging
\citep[e.g.][]{cimatti2008,vandokkum2008, toft2009, damjanov2011,
  szomoru2011}.  
With new, more sensitive NIR spectrographs like X-shooter, we are now
entering an exciting era where it is becoming possible to detect
absorption lines at sufficient S/N and resolution to constrain their
relative strengths, which are sensitive to the age and metallicity of
the stellar populations, but not to dust content, and their
broadening, from which we can derive the velocity dispersion and
dynamical mass.

In this paper we presented the first spectrum of a massive evolved
$z\sim2$ galaxy of sufficient S/N and resolution to perform detailed absorption
line spectroscopy. 
From the absorption lines we derive a velocity dispersion of
$\sigma_*=318\pm53 $km s$^{-1}$ and a dynamical mass of log(M$_{dyn}$/M$_{\odot}$)=11.51$\pm$0.11.
Simultaneous fits to absorption line indices and the full restframe
UV-optical spectrum allowed us, for the first time at $z\sim2$, to break the
degeneracies between age, dust and metallicity, resulting in more
realistic error bars on all derived quantities, including the stellar mass.
\begin{itemize}
\item
We derived a metallicty of
$\log(Z_\ast/Z_\odot)=0.02^{+0.2}_{-0.41}$,  which is comparable to that of similar mass
local galaxies.
\item We accurately determined the mass/luminosity weighted stellar age
  (to within 0.1 dex), which implies that the majority of the
stars formed in a burst at $z_f>3.3^{+0.5}_{-0.3}$ which continued for at least
another 0.4 Gyr (to $z\sim2.7$). The derived mean stellar age is
naturally much younger than in local galaxies, but if the galaxy
evolves passively to the present day, its stellar age will be similar
to local galaxies of similar mass.
\item
We independently confirmed the galaxies quiescent nature by deriving robust upper limits on the
specific starformation rate from the slope of the UV continuum and
upper limits on emission lines: sSFR $<4.4\times10^{-11}\rm{yr}^{-1}$.  The
shape of the continuum and the presence of strong Balmer absorption
lines resembles a mix of A and F stars, typical of post star burst
galaxies and consistent with the galaxy being quiescent, but with a significant
starburst relatively recently.

\end{itemize}

We compiled 3 additional spectroscopically confirmed massive quiescent $z\sim2$
galaxies with measured velocity
dispersion and dynamical masses from the literature, allowing for a
homogeneous study of scaling relations between structural and dynamical properties of
evolved galaxies at this high redshift. 

Two of the galaxies are also post starburst galaxies with similar ages
($\lesssim 1$Gyr), brightnesses
(L$_K$=15-23$L_K^{*}$), velocity dispersions, and sizes to the galaxy
analysed in this paper, while one
is less lumnious (L$_K$=7L$_K^*$) and much more compact. This galaxy
has a spectrum characteristic of more evolved stellar populations,
with less prominent Balmer lines, but stronger Ca H an K absorption
lines, and an estimated age of $>$ 1.5 Gyr. 
The $z\sim2$ post starburst galaxies are not as compact as the typical
quiescent $z\sim2$ galaxies studied previously in photometric samples.  
This may be a selection effect, as they were selected for spectroscopy
due to their brightness and therefore likely to be
biased towards younger, less evolved $z\sim2$ quiescent galaxies,
where the brightness have been boosted by a starburst in the not so
distant past. The evolved SEED MUSYC 1252-0 is an outlier in this
paper, but may be more representative of the quiescent $z\sim2$
galaxy population, with a fainter K-band magnitude, older
stellar population and higher stellar mass density. The sample allowed for:
\begin{itemize}
\item 
Studying the dynamical mass-size relation of quiescent galaxies at
$z\sim2$, which is independent on systematic uncertainties and
degeneracies introduced by photometric redshifts and stellar population synthesis modeling of
broad band photometry.
\item
The offset of the dynamical mass-size relation of the $z=2$ post
starburst galaxies from the local early type relation is smaller (a factor of 2.5) than the offset of the stellar
mass-size relation (a factor of 3.3), and its scatter appears to be
smaller (though this is hard to quantify, given the small sample).  
\item
The smaller offsets are caused by a difference in stellar to dynamical
mass ratios of the SEEDs and the local early type comparison
sample. Interpreting this as a difference in the central dark matter
fraction, the latter on average have a dark matter fraction of $0.46
\pm 0.23$ while the SEEDs on average have a dark matter fraction of
$0.18 \pm 0.20$ consistent with being completely baryon dominated.  
\end{itemize}
We also compared the $z\sim2$ galaxies to local post starburst
galaxies:
\begin{itemize}
\item
The local post starburst stellar mass-size relation is
similar to the local early type relation, but shifted along the
relation to higher masses and larger sizes.  The local post starburst
population show a shift between the stellar and dynamical mass-size
relation, similar to  the $z\sim2$ post starburst galaxies,
indicating a similarly low central dark matter fraction.
\end{itemize}
We compared the observed structural and dynamical properties of
our $z\sim2$ sample to predictions from the cosmological simulations
of \cite{oser2012} for the size, velocity dispersion and
dark matter fraction evolution of massive galaxies between $z=2$ and
$z=0$. 
The simulations do a qualitatively good job reproducing the difference
in the stellar mass-size relations, and the size-velocity dispersion
relations, and are also able to reproduce some (around 50\%) of the observed
difference in dark matter fraction between z=2 and 0, but not fully,
resulting in an overprediction of the observed evolution
in the dynamical mass-size relation.
To further compare the structural properties of the $z\sim2$ sample to
local galaxies, we constructed:
\begin{itemize}
\item
The first estimate of the fundamental
plane at this high redshift, which is offset from the
local fundamental plane by a larger amount than can be explained by 
passive evolution of the stellar populations, in other words they have
to go through structural evolution to evolve onto the fundamental
plane observed at lower redshift.
\end{itemize} 

The structural evolution implied by the
cosmological simulations brings the $z=2$ galaxies closer to the local FP, but still
significantly offset.
This suggests that additional, possibly non-homologous structural evolution
is needed. 
A possible mechanism responsible for this evolution is major mergers: Observations
show that most massive galaxies will undergo at least one major merger
between z=3 and 0 \citep[e.g.][]{man2012}, and while this may not be an
efficient mechanism for size growth \citep[e.g][]{bezanson2009}, it could increase
the central dark matter fraction \citep[through redistribution of the dark
matter,][]{boylan-kolchin2005} and provide some of the structural
evolution needed (e.g. changing the structure from disk like to bulge
like).

\begin{figure}
\begin{center}
\includegraphics[scale=0.6,angle=0]{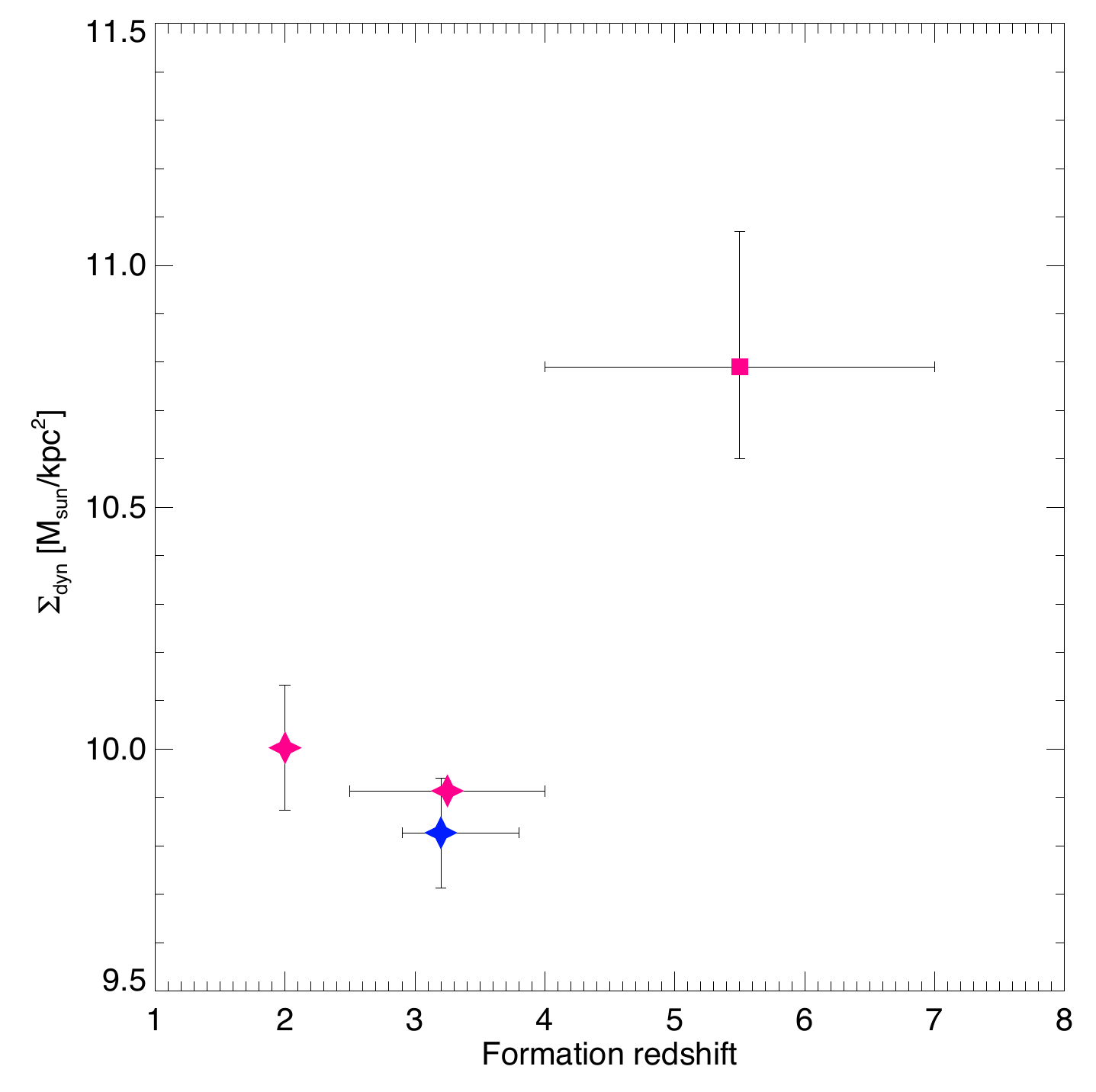}
\caption{Dynamical mass density versus the mean stellar formation redshift for
the $z\sim 2$ galaxies. The formation redshift is calculated from the
stellar ages, derived from the stellar population synthesis fits to
the spectra. The error bars on the mean formation redshifts of the three
galaxies from the literature are tentative, as they are not
accurately derived in the papers (in one case missing). This small
sample is consistent with a picture where galaxies formed at higher
redshift are denser.
 }
\end{center}
\label{fig.cosmicclocks}
\end{figure}

It has been suggested that SEEDs are more compact than galaxies of
similar mass assembled later in the history of the Universe, because
the Universe was much denser and more gas rich at earlier times
\citep[e.g.][]{khockfar2006}. If this is the case we may expect the
mass density of SEEDs to be larger for older, more evolved examples
than for younger ones. \cite{zirm2011} suggested that the peak stellar
density may scale with their formation redshift while the ratio of
high-density to low-density components would be related to the number
of 
(dry) mergers they have undergone. In Figure \ref{fig.cosmicclocks} we
plot the central (dynamical) mass density of the $z\sim2$ galaxies
versus their estimated formation redshift. Based on these four
galaxies alone it is not possible to determine if there is a relation,
but we note that the post starburst SEEDs have similar mass density
and formation redshift, which are both significantly smaller than in the
evolved SEED, consistent with this scenario.  The main uncertainty in
the plot is the lack of accurate ages/formation redshifts with
realistic error bars, for the three galaxies from the literature. The
papers describing the observations only give rough ranges of possible
ages for two of them, while for one (NMBS-C7447) no errors on the age
is quoted.  In a recent work on a photometric sample of quiescent
galaxies at $z\sim2$, \cite{szomoru2011} did not find a relation
between compactness and rest frame U-V color, which was used as a
proxy for stellar age.  However we can not rule out  that
this may be due to inaccuracy of the adopted age proxy.  As shown in
this paper, careful analysis of the absorption lines strength and the
continuum emission in X-shooter spectra allows for much more
accurate age estimates, making it possible to observationally test for
such a relation.  SEED galaxies are excellent targets for this
test. They are quiescent, with star formation history well fitted by
models with relatively well defined formation redshifts and they have
regular symmetric morphologies, making it relatively simple to
accurately determine their mass density.

In this paper we have presented homogeneous constraints on the $z\sim2$
dynamical mass-size relation, the first constraints on the fundamental plane and a possible
relation between stellar age and compactness. Relations that have until now
only been possible study out to $z\sim1.3$. In the immediate future it
will be possible to put such and related studies on secure statistical
footing as the samples of $z\sim2-3$ quiescent galaxies with
absorption lines spectroscopy increases. This will be a large leap
forward in our understanding of galaxy formation, as this is the era
where these and other scaling relations we observe at lower redshift
are believed to form.

\subsection{Acknowledgement}
We thank Lise Christensen, Tayyaba Zafar and Marijn Franx for sharing
their experience with X-shooter observations and data reduction, 
Ryan Quadri and Rik Williams for assistance with the target selection, and for providing their
UDS catalogs and images, Thorsten Naab for useful discussions related
to predictions from cosmological simulations, Jens Hjorth for carefull
reading of the manuscript and stimulating discussions, and Bettoni for sharing
his X-shooter observations of stars. We also thank the anonymous
referee for very useful suggestions.  We gratefully acknowledge support from the Lundbeck
foundation. The Dark Cosmology Centre is funded by the Danish National Research Foundation.

\appendix

\section{Stellar populations: comparison between different observational constraints}\label{appendix.stelpops}

As discussed in Section~\ref{sec.stellarpops}, we derive physical parameter estimates 
following a Bayesian approach in which we compare the observed stellar absorption 
features, the rest-frame UV and optical color with the predictions of a large 
library of stochastic SFHs. 
Taking advantage of the large wavelength coverage of the X-Shooter spectrum, in this Appendix 
we explore the sensitivity of the derived parameters and associated 
uncertainties to the
observational constraints adopted. In particular we repeat the
analysis by using either only one color (rest-frame optical or UV) in
addition to the absorption indices (entries 2 and 3 respectively in Table~\ref{tab.paramcompare}) 
or the pixel-by-pixel flux over the
rest-frame UV and optical range (up to 5900~\AA; entry 4 in Table~\ref{tab.paramcompare}). 
Finally, we compare with the results obtained by fitting only the absorption indices to
dust-free models and estimating the dust attenuation from the
difference between the observed J-H color and the color of the
redshifted, dust-free model spectra (thus effectively assuming a
single-screen dust distribution and an attenuation law
$A_\lambda\propto\lambda^{-0.7}$, and excluding models that would
predict a negative dust attenuation; entry 5 in Table~\ref{tab.paramcompare}). The latter approach 
is similar to the one adopted in the analysis of SDSS optical spectra by
  \cite{Kauffmann03} and \cite{Gallazzi05}. 
\begin{figure*}
\centerline{\includegraphics[width=18truecm]{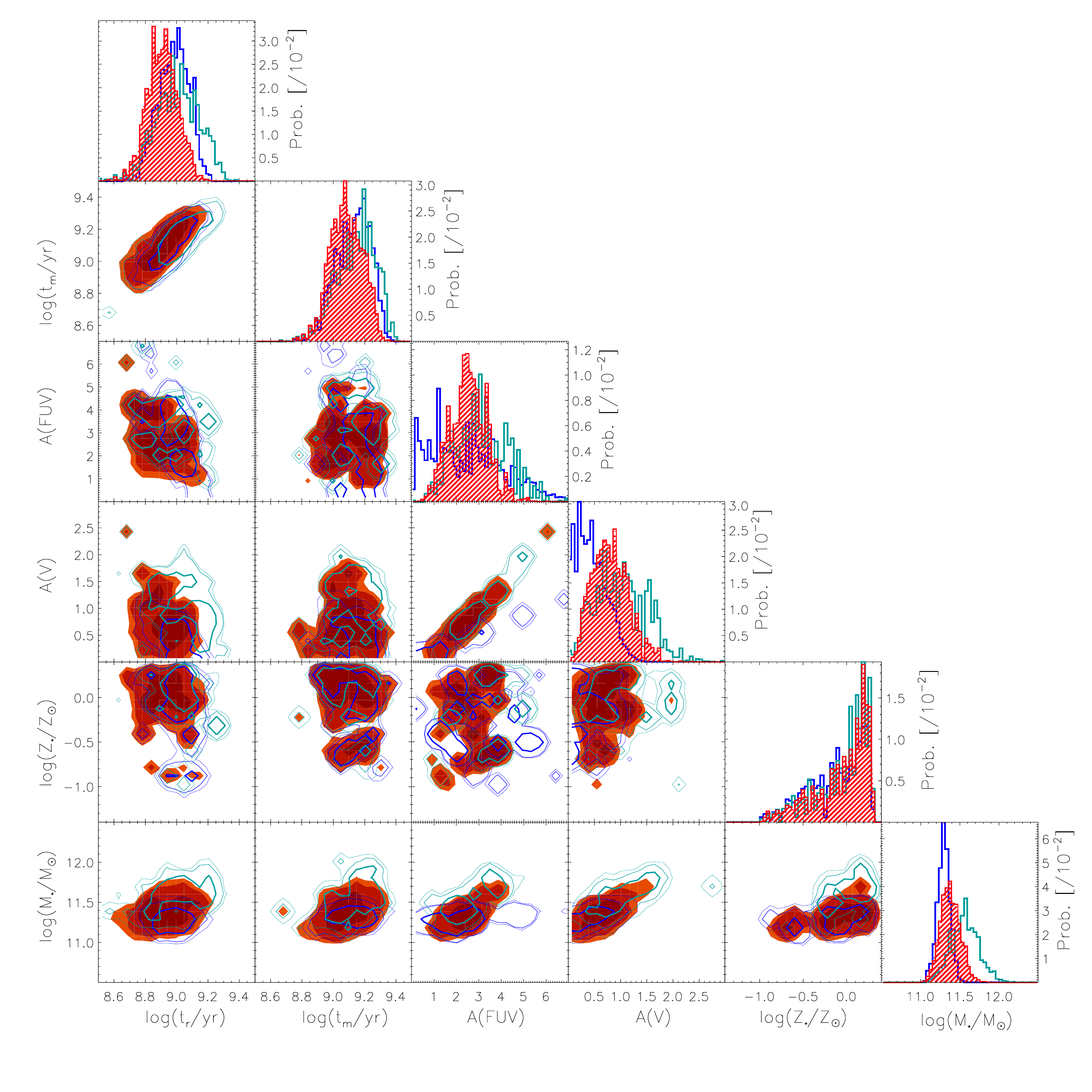}}
\caption{One- and two-dimensional marginalized Probability Density Functions for the physical parameters estimated using different 
observational constraints. The filled red histograms and contours show the results of our reference fit to the absorption indices, UV and 
optical colors. These are compared to the constraints obtained by fitting the absorption indices only (blue histograms and contours; fit 
5 in Table~\ref{tab.paramcompare}) or the pixel-by-pixel spectrum (green histograms and contours, fit 4 in Table~\ref{tab.paramcompare}). 
The contours enclose the 68\%, 85\% and 99\% confidence levels.
For clarity we omit the results obtained by fitting either the UV or the optical color in addition to the absorption indices, noting 
that they are bracketed by the PDFs shown here. }\label{fig.pdfs}
\end{figure*}

In Fig.~\ref{fig.pdfs} we show the full one- and two- dimensional 
Probability Density Functions of the physical parameters of interests (i.e. marginalized over 
all the parameters except the ones shown in each panel) as obtained with the different observational 
constraints. The results are summarized in terms of median and percentiles of the one-dimensional 
PDFs in Table~\ref{tab.paramcompare}. 
\begin{deluxetable}{lcccc}
\tablecaption{Stellar population parameter estimates and uncertainties
  obtained by fitting different observational constraints. The results 
  adopted in the paper are those obtained by simultaneously fitting 
 three stellar absorption features, the UV flux
ratio and the observer-frame J-H color (bold face). The other fits
explore the effect of changing the observational constraints (see text).}
\tablecolumns{5}
\tablewidth{0pc}
\tablehead{
\colhead{Method} & \colhead{$\log(t_r/yr)$} & \colhead{$\log(t_m/yr)$} & \colhead{$\log(Z_\ast/Z_\odot)$}  & \colhead{$\log(M_\ast/M_\odot)$}
}
\startdata
{\bf (1) indices, UV and J-H color} & $\mathbf{8.90^{+0.10}_{-0.09}}$ &
$\mathbf{9.08^{+0.11}_{-0.10}}$ & $\mathbf{0.02^{+0.20}_{-0.41}}$ & $\mathbf{11.37^{+0.13}_{-0.10}}$ \\
\noalign{\smallskip}
\hline
(2) indices, J-H & $8.94^{+0.10}_{-0.09}$ & $9.11^{+0.10}_{-0.11}$ & $-0.02^{+0.23}_{-0.45}$ & $11.30^{+0.12}_{-0.11}$ \\
\noalign{\smallskip}
(3) indices, UV flux ratio & $8.93^{+0.11}_{-0.10}$ & $9.10^{+0.11}_{-0.11}$ & $0.03^{+0.19}_{-0.41}$ & $11.47^{+0.20}_{-0.16}$ \\
\noalign{\smallskip}
(4) full SED & $9.01^{+0.14}_{-0.14}$ & $9.15^{+0.12}_{-0.14}$ & $0.04^{+0.20}_{-0.40}$ & $11.56^{+0.17}_{-0.19}$ \\
\noalign{\smallskip}
(5) indices, color excess & $8.99^{+0.10}_{-0.11}$ & $9.13^{+0.10}_{-0.13}$ & $-0.09^{+0.28}_{-0.38}$ & $11.29^{+0.08}_{-0.09}$ \\
\noalign{\smallskip}
%\enddata
\hline
\hline
\smallskip

\label{tab.paramcompare}
%\end{deluxetable}
%\begin{deluxetable}{lcccc}
%\tablecolumns{5}
%\tablewidth{0pt}
%\tablehead{

%\colhead{Method} & \colhead{$\tau_V$} & \colhead{$\mu\tau_V$} &
%\colhead{$A_{FUV}$} & \colhead{$A_V$}
%}
        & $\tau_V$ & $\mu\tau_V$ & $A_{FUV}$ & $A_V$ \\ \hline
%\startdata
{\bf (1) indices, UV and J-H color} & $\mathbf{1.78^{+0.88}_{-0.63}}$ & $\mathbf{0.64^{+0.32}_{-0.28}}$
&$\mathbf{2.52^{+0.89}_{-0.81}}$ &$\mathbf{0.77^{+0.36}_{-0.32}}$ \\
\noalign{\smallskip}
\hline
(2) indices, J-H & $1.37^{+0.81}_{-0.72}$ & $0.40^{+0.34}_{-0.23}$ &$1.83^{+0.93}_{-0.79}$ &$0.49^{+0.39}_{-0.28}$ \\
\noalign{\smallskip}
(3) indices, UV flux ratio & $2.04^{+1.01}_{-0.66}$ & $0.83^{+0.45}_{-0.36}$ &$3.04^{+1.19}_{-0.98}$ &$0.98^{+0.50}_{-0.41}$ \\
\noalign{\smallskip}
(4) full SED & $2.10^{+1.024}_{-0.74}$ & $0.85^{+0.46}_{-0.43}$ &$3.09^{+1.25}_{-1.14}$ &$0.98^{+0.53}_{-0.48}$ \\
(5) indices, color excess & $--$ & $--$ &$2.31^{+1.93}_{-1.78}$ &$0.39^{+0.33}_{-0.31}$ \\
\noalign{\smallskip}
\enddata
\end{deluxetable}

\begin{figure*}
\centerline{\includegraphics[width=18truecm]{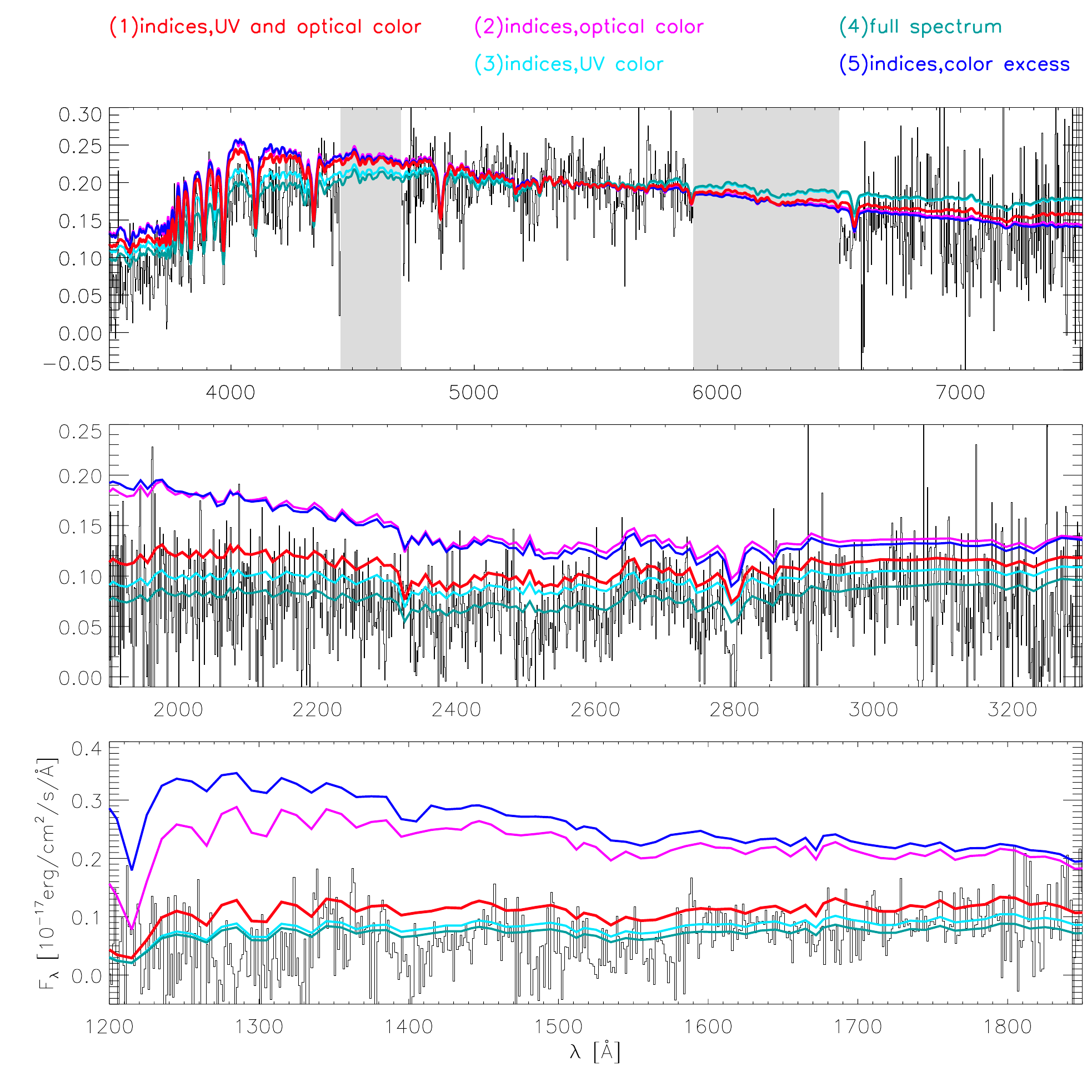}}
\caption{X-Shooter spectrum at medium-resolution (3{\AA}/bin in the NIR arm, 16{\AA} per bin in the UVB and VIS arms) over the
full wavelength range covered by the three arms. Overplotted are the bestfit model spectra obtained by constraining {\bf{1)}} three
stellar absorption indices, one rest-frame optical color and the UV flux ratio (red, our default fit), {\bf{2)}} three stellar
absorption indices and one rest-frame optical color (magenta), {\bf{3)}} three stellar absorption indices and the UV
flux ratio (cyan), {\bf{4)}} the full spectrum pixel-by-pixel up 5900{\AA} (green), and {\bf{5)}} three stellar absorption indices and a cut on
color excess (blue). Fits obtained by constraining the optical wavelength range only under-predict the
dust attenuation in the UV and hence over-predict the UV flux. In the NIR spectrum, we have masked out areas of high background contamination.}
\label{fig.dust}
\end{figure*}

We note the following:\\
-- The luminosity- and the mass-weighted
age estimates agree well within the typical uncertainty of
$\sim$0.1~dex, although the pixel-by-pixel fit tends to provide
slightly older ages than the fits to the absorption indices alone or
in combination with broad-band information;\\
-- There is a mild age-dust degeneracy in the sense that older ages are 
associated with lower dust attenuation. Instead, there is no clear 
age-metallicity degeneracy, partly because of the different sensitivity of 
the absortion indices to the two parameters, and partly because of the  
larger uncertainties in metallicity with respect to age as expected for young stellar 
populations \citep[e.g.][]{Gallazzi05}. In general degeneracies between parameters are attenuated when fitting individual 
absortion features rather than the full SED; \\ 
-- The different slope in the A(FUV)--A(V) PDFs is a consequence of the different dust 
geometry assumed;\\
-- All the fits indicate a rather large attenuation by dust, although in general fits
to the rest-frame optical wavelength range tend to predict lower
overall dust attenuation, and in particular under-estimate the
attenuation in the UV. The need to include constraints from the UV in order to correctly predict the dust attenuation 
is further illustrated in Fig.~\ref{fig.dust}
where the bestfit models to the five different sets of observational constraints are over-plotted on
the observed spectrum: the best fits to the optical over-predict the
flux in the UV.\\
Overall there is good agreement among the various fits, in particular on stellar age. We note 
that our default fit combines the sensitivity of UV and optical colors to dust attenuation 
and the ability of stellar absorption features to alleviate parameter degeneracies.

\section{Fitting X-shooter Template Stellar Spectra}
\label{appendix.vdisp}
\begin{figure}[htbp]
\begin{center}
\includegraphics[scale=0.7,angle=0]{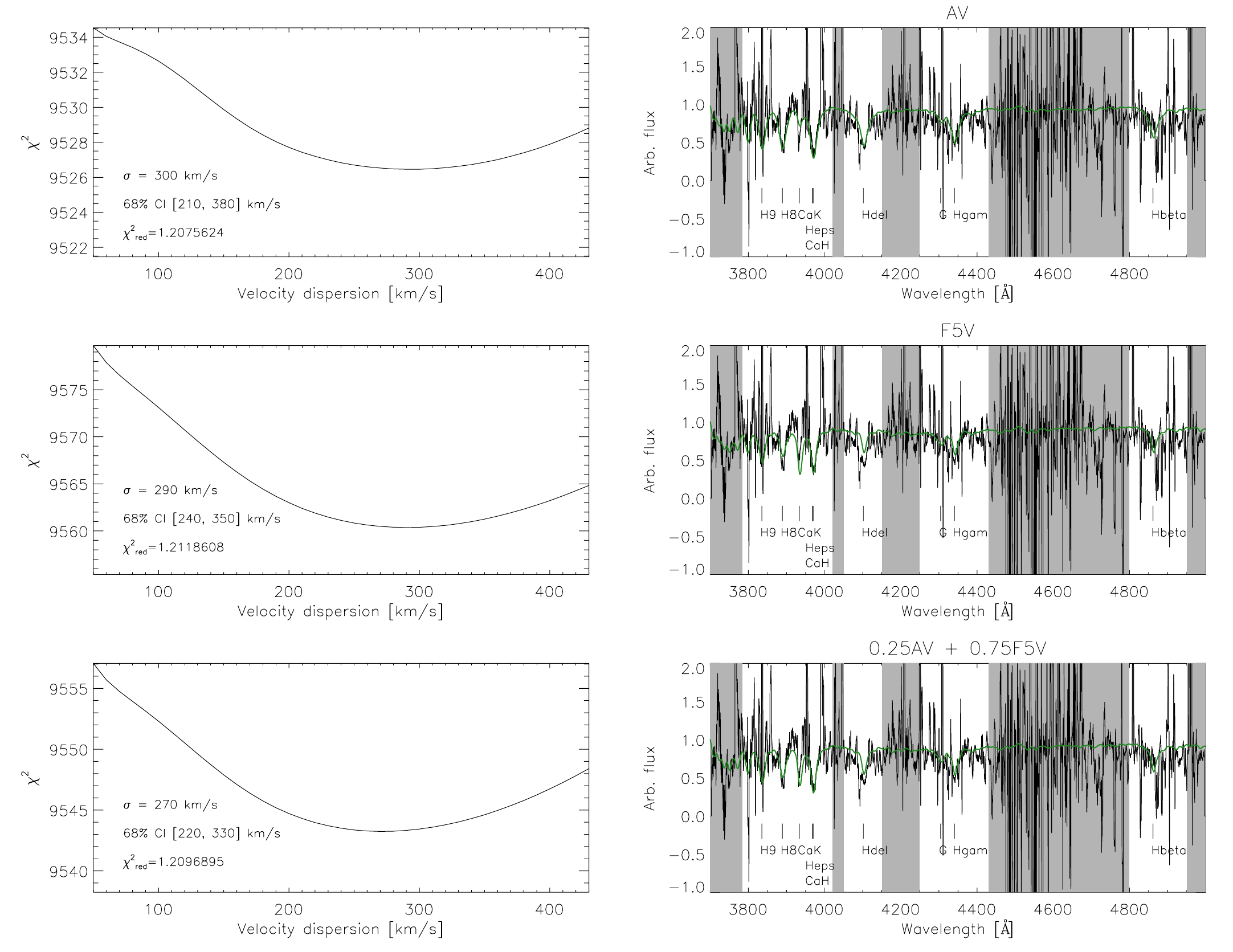}
\caption{Fits to the galaxy spectrum with the A and  F star, and 
  the best fitting mixture. Regions excluded in the fit are shaded in grey. The galaxy
  spectrum is smoothed to a resolution of 10 {\AA} for displaying purposes.}
\label{figure:finalfig}
\end{center}
\end{figure}
As an independent test of the velocity dispersion derived with pPXF
we fit stellar template spectra convolved with Gaussians to the galaxy
spectrum to estimate the velocity dispersion through a chi-sqared
analysis \citep[see][]{barth2002, wold2007}.  For stellar templates
we use stars of five different spectral types (AV, F5V, G8IV, K0III,
K0V), observed with X-shooter at the same slit width, for a different
program (084.B-035A, PI: Hjorth).  In order to estimate the velocity
dispersion of the galaxy we convolved these stellar spectra with
Gaussians and fitted them to the Balmer line region of the galaxy
spectrum

The rest-frame wavelength range of the NIR arm in the galaxy spectrum
corresponds approximately to the wavelength range of the UVB arm in
the stellar spectra. When we do the spectral fitting, we therefore
compare the stellar UVB to the NIR galaxy spectrum.  The stellar
spectra were reduced with a spectral sampling of 0.2 {\AA}/pix in the
UVB arm, and 0.5 {\AA/pix in the NIR arm. The spectral sampling in the
  NIR arm of the galaxy spectrum corresponds to 0.16 {\AA}/pix in the
  galaxy rest-frame. As this is very similar to the sampling of the
  UVB stellar spectra, and because of the lower quality of the galaxy
  spectrum compared to the stellar spectra, we chose to use the
  spectra as they are without further manipulation and resampling.
  Both the galaxy spectrum and the stellar spectra were normalized by
  their continua before fitting. For the stellar spectra we extracted
  the continuum by marking anchor points in the spectra and doing a
  spline fit with varying degrees of tension.
  Due to the lower S/N we used a different approach for the galaxy
  spectrum, and extracted the continuum shape from the best fitting
  model of the galaxy (see Section \ref{sec.stellarpops}) with
  sampling every 50 {\AA}.
\begin{deluxetable}{ccccc}
\tablecolumns{5}
\tablecaption{Stellar templates used in velocity dispersion fitting. The best fit velocity dispersion and 68 per
cent confidence interval is listed in the last two columns. All fits
were done in the region between 3785 and
5000 {\AA} (restframe) as explained in the text}
\tablewidth{0pt}
\tablehead{
\colhead{Template} & \multicolumn{2}{c} {Spectral type mix (\%)}  & \colhead{$\sigma$} & \colhead{68\% CI} \\
  & \colhead{AV} & \colhead{F5V}  & \colhead{[km\,s$^{-1}$]}& \colhead{[km\,s$^{-1}$]}
}
\startdata
1 & 100 & 0  & 300 & [210,380] \\ 
2 & 0 & 100  & 290 & [240,350] \\
%3&  0 & 0     & 100  & 330 & [280,400] \\
4 & 75 & 25   & 270 & [200,340] \\
5 &  50 & 50  & 260 & [200,330] \\ 
6 & 25 & 75  & 270 & [220,330]\\
\enddata
\label{tab.templatelist}
\end{deluxetable}
We experimented with fitting different regions of the spectrum to
maximize the S/N in the velocity dispersion measurement, including as
many absorption lines as possible, while down-weighing regions of the
spectrum with low S/N or strong skyline residuals.  We obtained the
best results by fitting the region between rest-frame wavelength 3785
{\AA} and 5000 {\AA} where most of the strong absorption features are
found. This region covers the region bluewards of $H{\beta}$. The MgI
triplet at $\approx$5175 {\AA} redwards of H$\beta$ and the Na doublet
(5889.95, 5895.92 {\AA}) are not detected in the spectrum, hence we
concentrated on estimating the velocity dispersion from the Balmer
line region.
The AV and F5V stars (which have prominent Balmer lines) provide good
fits with well constrained minima in the $\chi^2$ distribution. The
other stars provide significantly worse fits. The K0V has a minimum
consistent with the best fitting values of the K0V and F5V star, but
with a much flatter $\chi^2$ distribution, while the G8IV and K0III do
not produce a minimum, due to too weak Balmer lines, and too strong
G-band and CaH+K lines.  We also fit different linear combinations of
the AV and F5V stars. These fits produce consistent best fitting
values, but with slightly smaller errorbars.  In Figure
\ref{figure:finalfig} and Table \ref{tab.templatelist} we show
the results from the best fits. From this analysis the best fitting
velocity dispersion is in the range $260-300$ km s$^{-1}$ with a typical
(random) uncertainty of $50 $km s$^{-1}$, in agreement with the
results from the pPXF fits.

\end{document}